\documentclass[floatfix,footinbib,reprint,onecolumn,prb,aps,a4paper,preprintnumbers,amsmath,amssymb,longbibliography]{revtex4-1}
\usepackage{graphicx}
\usepackage[colorlinks=true,linkcolor=blue,citecolor=blue,urlcolor=blue]{hyperref}
\usepackage{upgreek}
\usepackage{soul}

\usepackage{bm}
\usepackage{multirow}
\usepackage{siunitx}

\usepackage{amsfonts}
\usepackage{color}
\usepackage{mathtools}
\newcommand*\diff{\mathop{}\!\mathrm{d}}
\newcommand{\trace}{\text{Tr}}
\newcommand{\iu}{\mathrm{i}}
\newcommand{\mat}[1]{\underline{\underline{#1}}}

\usepackage{ragged2e}

\makeatother

\begin{document}
\title[Multiple-scattering approach for multi-spin chiral magnetic interactions]{Multiple-scattering approach for multi-spin chiral magnetic interactions: Application to one- and two-dimensional Rashba electron gas}
\author{Samir Lounis}\email{s.lounis@fz-juelich.de}

\address{ Peter Gr\"{u}nberg Institut and Institute for Advanced Simulation,
	Forschungszetrum J\"{u}lich and JARA, 52425 J\"{u}lich, Germany}	

\begin{abstract}
Various multi-spin magnetic exchange interactions (MEI) of chiral nature have been recently unveiled. Owing to their potential impact on the realisation of twisted spin-textures, their future implication in spintronics or quantum computing is very promising.  Here, I address the long-range behavior of multi-spin MEI on the basis of a multiple-scattering formalism implementable in 
 Green functions based methods such as the Korringa-Kohn-Rostoker Green function framework. I consider the impact of spin-orbit coupling (SOC) as described in the one- (1D) and two-dimensional (2D) Rashba model, from which the analytical forms of the four- and six-spin interactions are extracted and compared to  the well known bilinear isotropic, anisotropic and Dzyaloshinskii-Moriya interactions (DMI). Similarly to the DMI between two sites $i$ and $j$, there is a four-spin chiral vector perpendicular to the bond connecting the two sites. The oscillatory behavior of the MEI and their decay as function of interatomic distances are analysed and quantified for the Rashba surfaces states characterizing Au surfaces. The interplay of beating effects and strength of SOC gives rise to a wide parameter space where chiral MEI are more prominent than the isotropic ones. The multi-spin interactions for a plaquette of $N$ magnetic moments decay like $\{q_F^{N-d}  P^{\frac{1}{2}(d-1)}L\}^{-1}$  simplifying to $\{q_F^{N-d}  R^{\left[1+\frac{N}{2}(d-1)\right]}N\}^{-1}$ for equidistant atoms, where $d$ is the dimension of the mediating electrons, $q_F$ the Fermi wave vector, $L$ the perimeter of the plaquette while $P$ is the product of interatomic distances. This recovers the behavior of the bilinear MEI, $\{q_F^{2-d} R^{d}\}^{-1}$, and shows that increasing the perimeter of the plaquette weakens the MEI. 
More important, the power-law pertaining to the distance-dependent 1D MEI is insensitive to  the number of atoms in the plaquette in contrast to the linear dependence associated with the 2D MEI. Furthermore, the $N$-dependence of $q_F$ offers the possibility of tuning the interactions amplitude by engineering the electronic occupation. 
\end{abstract}

\maketitle

\section{Introduction}
Non-collinear spin-textures, such as the one-dimensional (1D) domain walls~\cite{Parkin1990}, 2D  magnetic skyrmions~\cite{Fert2017} and 3D hedgehog states~\cite{kanazawa_critical_2016,Bornemann2019} or magnetic  hopfions~\cite{Wang2019} are at the heart of investigations exploring their potential realization as magnetic bits for future information technology. 
Likewise, such complex spin-states are key players in quantum computing by being  paramount in the physics of non-Abelian Majorana bound states~\cite{Read2000,Nayak2008,Nadj-Perge2013,Klinovaja2013,Braunecker2013,Vazifeh2013,Kim2014,Feldman2017,Kim2018,Schneider2020} induced by magnetic nanostructures at the vicinity of superconductors and in several proposals for topological superconductivity~\cite{Fu2008,Sau2010,Alicea2010,Lutchyn2010,Oreg2010}.

The occurrence of complex magnetism is driven by the existence of various competing magnetic exchange interactions (MEI). The initially proposed interactions are bilinear in nature such as the isotropic MEI, $J^\text{iso}_{ij}\, (\mathbf{S}_i\cdot\mathbf{S}_j)$\cite{Heisenberg1928,Anderson1959}, involving two spins while the Dzyaloshinskii-Moriya interaction (DMI), $\mathbf{D}_{ij}\cdot (\mathbf{S}_i\times\mathbf{S}_j)$, requires the presence of spin-orbit coupling (SOC) and broken inversion symmetry\cite{Dzyaloshinskii1958,Moriya1960}. The direction of the DM vector,  $\mathbf{D}$, settles the sense of rotation of the magnetic moments, defining thereby and usually the chirality of the magnetic structures. Besides the isotropic MEI and DMI, a compass term, $J^\text{ani}$\cite{vanVleck1937,Moriya1953}, can be finite, which define the tensor of bilinear (two-site two-spin) MEI. $J^\text{ani} \in\mathbb{R}^{3\times 3}$ represents a (trace-less symmetric) matrix in the space of spin-moment orientation 
giving rise to a non-local magnetic anisotropy energy and behaves like the latter quantity.

The importance of going beyond the bilinear magnetic interactions was recognized several decades ago. Solid $^3$He is a good example, where multi-spin isotropic interactions are a key ingredient to grasp its magnetic behavior~\cite{Roger1983,Roger1998}.  
It is established that if not properly taken into account, such higher-order interactions contribute to effective bilinear ones (see e.g. Refs.~\cite{Lounis2010,Szilva2013}). In fact, these interactions can involve an unequal number of magnetic moments and number of sites\cite{Al-Zubi2011,Kroenlein2018,Romming2018} but without breaking time reversal symmetry, the number of magnetic moments should be even.  For instance, the biquadratic interaction,  connecting four spin positioned on two sites, can be isotropic~\cite{Kittel1960,Harris1963,Huang1964}, $\mathcal{B}_{ij} (\mathbf{S}_i\cdot\mathbf{S}_j)^2$, or chiral as recently discovered~\cite{Brinker2019}, $\boldsymbol{\mathcal{C}}_{ij}\cdot(\mathbf{S}_i\times\mathbf{S}_j) (\mathbf{S}_i\cdot\mathbf{S}_j)$. The latter follows Moriya's rules similarly to the DMI. Generalization to four-spin interactions can be performed at the isotropic~\cite{Uryu1965,Iwashita1974,Hoffmann2020} 
or the chiral level~\cite{Brinker2019,Laszloffy2019,Grytsiuk2020,Brinker2020}, which are found to be large with non-negligible impact on the magnetic behavior of materials~\cite{Brinker2019,Grytsiuk2020,Brinker2020}. The microscopic origin of the chiral four-spin interactions was given in Ref.~\cite{Brinker2019}, where the focus was on terms proportional to 
$\boldsymbol{\mathcal{C}}_{ijkl}\cdot(\mathbf{S}_i\times\mathbf{S}_j) (\mathbf{S}_k\cdot\mathbf{S}_l)$, while those of the form $(\boldsymbol{\mathcal{C}}_{ijkl}\cdot\mathbf{S}_i) \chi_{jkl}$, with $\chi_{jkl} = \mathbf{S}_j\cdot (\mathbf{S}_k\times\mathbf{S}_l)$ being the scalar three-spin chirality, were addressed and derived in Ref.~\cite{Grytsiuk2020} after a systematic 
 multiple-scattering expansion in the spirit of the infinitesimal rotation method of the magnetic moments. The scalar three-spin chirality is finite whenever the three magnetic moments are non-coplanar. In such non-coplanar spin-configurations, an orbital moment, chiral in nature, can emerge even without SOC~\cite{Hoffmann2015,Dias2016,Hanke2016,Dias2017}. This implies that the chiral four-spin interaction can be interpreted as an interaction between the spin and chiral orbital moments.   
 One notes that subsequently a derivation of some of the four-spin terms has been proposed in~\cite{Mankovsky2019}. As shown in  Ref.~\cite{Brinker2020}, the various forms of the chiral four-spin interactions can be re-expressed in terms of sums involving $\boldsymbol{\mathcal{C}}_{ijkl}\cdot(\mathbf{S}_i\times\mathbf{S}_j) (\mathbf{S}_k\cdot\mathbf{S}_l)$. Also it is demonstrated that these interactions do not have to follow Moriya's rules of the DMI. Surprisingly, even isotropic six-spin interactions, which can be written in terms of a product of scalar three-spin chiralities $\chi_{ijk}\chi_{lmn}$, can be considerably large  as shown by Grytsiuk et al.~\cite{Grytsiuk2020} for B20-MnGe.

The recent developments triggered by the investigations  of  multi-spin interactions motivate the present study. Their potential impact on the magnetism of materials and their magnetodynamical properties remains to be explored.
The  focus here is on the  long-range behavior of such interactions, which is textbook knowledge for  the isotropic bilinear Ruderman-Kittel-Kasuya-Yosida (RKKY) interactions ~\cite{Ruderman1954,Kasuya1956,Yosida1957}. The latter have key implications in the interlayer exchange coupling of multilayers\cite{Gruenberg1986,Carbone1987,Parkin1990,Bruno1991} and when taken to their isotropic four-spin form, they can trigger  the superposition of multiple $Q$-states~\cite{Batista2016,Ozawa2017,Hayami2017,Okumura2020} and the stabilization of 2D magnetic skyrmions~\cite{Heinze2011} or possibly 3D magnetic textures~\cite{Tanigaki2015,Takagieaau2018,Fujishiro2019,Kurumaji2019,Khanh2020}. In the context of nuclear spins embedded in a Luttinger liquid, the RKKY interactions emerging from weak hyperfine interaction between the nuclear spin and the conduction electron spin can trigger complex helimagnetic phases. A strong renormalization of the nuclear Overhauser field is predicted accompanied with a universal reduction in the electric conductance and a potential gaping of the electronic spectrum~\cite{Braunecker2009,Meng2013}.

Note that long-range interactions are intimately connected to conventional and complex magnetic Friedel oscillations~\cite{Friedel1952,Walls2007,Lounis2012}, which do not only affect the isotropic MEI, the DMI~\cite{Smith1976,Fert1980,Imamura2004,Khajetoorians2016,Bouaziz2017,Han2019,Schmitt2019} 
but also the magnetic anisotropy energy~\cite{Khajetoorians2013,Bouhassoune2016}, the magnetic stability and spin-fluctuations of  nanostructures~\cite{Hermenau2019}. The goal of this article is to establish and analyse the long-range behavior of chiral multi-spin interactions within the Rashba model~\cite{Rashba1960,Bychkov1984} used to describe the 1D and 2D electron gas subjected to the Rashba spin-orbit coupling. For that a multiple-scattering approach based on Green functions partly presented in Ref.~\cite{Grytsiuk2020} is utilized after reviewing it. The multi-spin interactions are quantified on the basis of parameters describing the Rashba 1D and 2D surface states characterizing Au surfaces~\cite{Reinert2003,Mugarza2001,Ortega2002}. I consider localized magnetic atomic spins, which can be manipulated and probed on surfaces utilizing scanning tunneling microscopy/spectroscopy (STM/STS). This tool became ubiquitous to quantify the isotropic\cite{Wahl2007,Meier:2008,Zhou2010,Meier2011,Khajetoorians2012,Prueser2014} and DM\cite{Khajetoorians2016} long-range bilinear interactions. In some cases, puzzling magnetic responses were probed in exotic man-crafted structures\cite{Khajetoorians2012} indicating that 
there is still a lot to unravel in this field of research. STM/STS would be an ideal tool to probe the multi-spin interactions addressed in this work. 

\section{Multiple-scattering expansion of the magnetic interactions}

The extraction of the MEI characterizing a magnetic material can be made via the celebrated infinitesimal rotation method of the magnetic moments~\cite{Liechtenstein1987,Udvardi2003,Ebert2009}, which defines a mapping procedure between the energy obtained from electronic structure calculations and the energy of an extended Heisenberg model. The magnetic force theorem~\cite{Heine1980,Oswald1985} permits to use the band energy instead of the total energy to 
evaluate the impact of rotating magnetic moments. The latter modifies the atomic potential by $\delta V$, which alters the density of states $\delta n$ and the corresponding energy
\begin{eqnarray}
\delta E &=& \int_{-\infty}^{\epsilon_F} \diff\epsilon\,(\epsilon-\epsilon_F)\delta n(\epsilon) = -\int_{-\infty}^{\epsilon_F} \diff\epsilon\,\delta N(\epsilon),
\end{eqnarray}
where $\delta N (\epsilon) =\int_{-\infty}^{\epsilon_F}  \diff\epsilon\,\delta n(\epsilon)$ represents the change of the integrated local density of states up to the Fermi energy $\epsilon_F$.

Since the density of states can be obtained from the imaginary part of the one-electron retarded  Green function ${G}(\epsilon) = (\epsilon -{H})^{-1} $ corresponding to the hamiltonian ${H}$ of the system: 
\begin{eqnarray}
n(\epsilon) &=& -\frac{1}{\pi}\mathrm{Im}\int_{\Omega_i} d\mathbf{r} \,\trace_s \sum_i G_{ii}(\mathbf{r},\mathbf{r},\epsilon),
\end{eqnarray}
where the integration is performed over the volume $\Omega_i$ defining the atomic site $i$ and the trace is taken over the spin index $s$, one can use  multiple-scattering theory as implemented in the Korringa-Kohn-Rostoker (KKR) Green-function method~\cite{Papanikolaou2002} to evaluate $\delta n$,  $\delta N$ and therefore $\delta E$. Because of the perturbation $\delta V$, the new Green function $g(\epsilon) = (\epsilon - H -\delta V)$, can be calculated from the original Green function G using the Dyson equation: $g(\epsilon) = G(\epsilon) + G(\epsilon) \delta V g(\epsilon)$.  The change in the Green function
\begin{eqnarray}
\delta g(\epsilon)=g(\epsilon) - G(\epsilon)=G(\epsilon)\delta V G(\epsilon)(1-G(\epsilon)\delta V)^{-1}
\end{eqnarray}

combined with the relation $\frac{dG}{dE}= -GG $ permits to find $\delta N(\epsilon) = -\frac{1}{\pi} \mathrm{Tr}\, \mathrm{ln}
[1 - G(\epsilon)\delta V]$, where the trace is taken over the position $\mathbf{r}$, site $i$ and spin index. Consequently, after perturbation, the modified energy reads
\begin{eqnarray}
\delta E = \frac{1}{\pi}\mathrm{Im} \int_{-\infty}^{\epsilon_F}\diff\epsilon\,\mathrm{Tr}\, \mathrm{ln}
[1 - G(\epsilon)\delta V]\,,
\end{eqnarray}
known as the Lloyd's formula~\cite{Lloyd1972,Drittler1989},   leading to a power series expansion
\begin{eqnarray}
\delta E = -\frac{1}{\pi}\mathrm{Im} \int_{-\infty}^{\epsilon_F} \diff\epsilon\,
 \sum_p \frac{1}{p}\mathrm{Tr}\, [G(\epsilon)\delta V]^p\,
 \label{Lloyd1}
\end{eqnarray}
  which is given in a matrix notation. 
More details on the derivation can be found for example in Refs.~\cite{Liechtenstein1987,Udvardi2003,Ebert2009,Lounis2010}.

In practice, the evaluation of the inter-site interactions requires the off-site part of the Green function instead of the full one. The off-site elements can be expanded as 
\begin{eqnarray}
G_{ij,ss'}(\mathbf{r},\mathbf{r}';\epsilon)=\sum_{LL'} R^\times_{iLs}(\boldsymbol{r};\epsilon) G^\mathrm{str}_{ij,LL',ss'}(\epsilon) R_{jL's'}(\boldsymbol{r'};\epsilon) 
\end{eqnarray}
within multiple scattering theory as implemented in KKR. $G^\mathrm{str}$ is the  structural Green function, which in contrast to the full Green function has no radial dependence but is  site-dependent. It is also a matrix in angular orbital momentum $L=(l,m)$ and spin representation,  $L\oplus s$. $R$ and $R^\times$ are respectively the left- and right-hand wave functions resulting from the potential $V$ without the perturbation. They are scattering solutions of either the Dirac or of the Schr\"odinger/scalar-relativistic equations augmented with the SOC. This allows one to utilize  the single-site scattering $t$-matrix instead of the radial-depenent atomic potential $V$ when evaluating the modified energy:
\begin{eqnarray}
\delta E  
&=& - \frac{1}{\pi}\mathrm{Im}  \int_{-\infty}^{\epsilon_F} \diff\epsilon\,\sum_p \frac{1}{p}\trace
[{G^{\mathrm{str}}}(\epsilon) \delta {t}(\epsilon)]^p\,,
\label{Lloyd}
\end{eqnarray}
where the trace is taken this time over the sites  $i$, orbital momentum  and spin indices. Similarly to the structural Green function ${G}^{\mathrm{str}}$,  $\delta {t}$ is a matrix in angular momentum and spin representation, $L\oplus s$. Eq.~\ref{Lloyd} is in practice  more convenient than Eq.~\ref{Lloyd1} since the latter hinges on cumbersome radially dependent quantities and radial integrations.

Adopting the rigid spin approximation implies a change of the magnetic part of the single-site $t$-matrix, which can be expressed as 
\begin{eqnarray}
\delta {t}_i (\epsilon)= {t}^{{\sigma}}_i(\epsilon) \,\delta \mathbf{S}_i  \cdot \boldsymbol{\sigma}\,.
\end{eqnarray}

Similarly to the proposal made in Ref.~\cite{Ebert2009}, ${t}^{\mathrm{s}}_i(\epsilon)$ describes the scattering at the magnetic part of the atomic potential or of the exchange and correlation potential, $B_{xc}$, if one works in the framework of density functional theory: $
    {t}^{\sigma}_{\substack{i,LL',ss'}}(\epsilon) = \int_{\Omega_i} d\boldsymbol{r} R_{iLs}^\times(\boldsymbol{r};\epsilon) B_{xc}(\boldsymbol{r}) R_{iL's'}(\boldsymbol{r};\epsilon)\,$.

The Green function can be decomposed into two Green functions, $A$ that is non-magnetic and diagonal in spin space and $\mathbf{B}$ that contains the magnetic part and a contribution induced by the spin-orbit coupling (SOC):
\begin{eqnarray}
{G}^\mathrm{str}_{ij} = A_{ij}\,\sigma_{0} 
+ \mathbf{B}_{ij}\cdot\boldsymbol{\sigma}\, ,
\label{Green_decomposed}
\end{eqnarray}
where $\boldsymbol{\sigma}$ is the vector of Pauli matrices and $\sigma_{0} $ is the identity matrix. If inversion symmetry is not broken, $A_{ij}$ is symmetric with respect to site exchange and likewise for 
${\mathbf{B}}^\sigma_{ij}$ if SOC is not present. ${\mathbf{B}}^\sigma_{ij}$ could have a magnetic contribution, which should behave like $A_{ij}$, and a contribution from SOC. Within the Rashba model, where for example broken-inversion symmetry and SOC are incorporated, ${\mathbf{B}}^\sigma_{ij} = -{\mathbf{B}}^\sigma_{ji}$. Note that the decomposition of the Green functions as proposed by Cardias et al.~\cite{Cardias2020} is useful for the investigation of multi-spin excitations.

\subsection{Two-spin interaction terms}

Considering infinitesimal rotations, I proceed to a Taylor expansion of the logarithm in Eq.~\eqref{Lloyd} and extract  systematically high-order terms ordered according to powers of  $\delta {t}$. As aforementioned terms with odd powers of $\delta {t}$ cancel since they are not compatible with the time-reversal symmetry requirement of the total energy. The standard second order term 
\begin{align}
 \delta E_{2\text{-}\mathrm{spin}} &= - \frac{1}{2\pi}  \mathrm{Im}\,\trace
 \int_{-\infty}^{\epsilon_F} \diff\epsilon\,  \sum_{ij}  {G}^\mathrm{str}_{ij}(\epsilon) \delta {t}_j(\epsilon){G}_{ji}^\mathrm{str}(\epsilon) \delta {t}_i(\epsilon) \label{eq:energy_2spin}
 \\&= -
 \sum_{ij}\left[J^{\mathrm{iso}}_{ij} \delta \mathbf{S}_{i} \cdot \delta\mathbf{S}_{j} + \delta\mathbf{S}_{i}  \cdot \mat{J}^{\mathrm{ani}}{}_{\mkern -22mu ij}\cdot\delta\mathbf{S}_{j} + \mathbf{D}_{ij} \cdot \left(\delta\mathbf{S}_{i}\times \delta\mathbf{S}_{j}\right)\right]
 \end{align}

 The integrand of Eq.~\ref{eq:energy_2spin} can be rewritten as
 \begin{align}
 \mathrm{Tr}\,{G}^\mathrm{str}_{ij}(\epsilon) \delta {t}_j(\epsilon){G}_{ji}^\mathrm{str}(\epsilon) \delta {t}_i(\epsilon) &= \mathrm{Tr}\,[A_{ij}\,\sigma_{0} 
+ \mathbf{B}_{ij}\cdot\boldsymbol{\sigma}]\, {t}^{\sigma}_j(\epsilon) \,\delta \mathbf{S}_j  \cdot \boldsymbol{\sigma} \,[A_{ji}\,\sigma_{0} 
+ \mathbf{B}_{ji}\cdot\boldsymbol{\sigma}]\, {t}^{\sigma}_i(\epsilon) \,\delta \mathbf{S}_i  \cdot \boldsymbol{\sigma}  \,.
 \end{align}

 Here one can use known relations involving products with Pauli matrices: $ (\mathbf{a} \cdot \boldsymbol{\sigma}) (\mathbf{b} \cdot \boldsymbol{\sigma}) = (\mathbf{a}\cdot\mathbf{b})\, \sigma_0 + i (\mathbf{a}\times \mathbf{b}) \cdot \boldsymbol{\sigma}$ to recover the  interaction energy between two magnetic moments from which the bilinear MEI can be extracted: 
\begin{align}
J_{ij}^{\mathrm{iso}} &=  \frac{1}{\pi} \mathrm{Im}\, \trace
\int_{-\infty}^{\epsilon_F} \diff\epsilon  
\mathcal{J}_{ij}^{\mathrm{iso}} (\epsilon)
\,,\label{eq_bilinear1}
\\
\mathbf{D}_{ij} &=   \frac{1}{\pi} \mathrm{Re} \, \trace \int_{-\infty}^{\epsilon_F} \diff\epsilon\, 
\boldsymbol{\mathcal{D}}_{ij}(\epsilon) \,\,\,\, \label{eq_bilinear2}
\\
\noalign{\noindent \vspace{2\jot}}
\delta\mathbf{S}_{i}  \cdot \mat{J}^{\mathrm{ani}}{}_{\mkern -22mu ij}\cdot\delta\mathbf{S}_{j} &=  \frac{1}{\pi} \mathrm{Im}\, \trace
\int_{-\infty}^{\epsilon_F} \diff\epsilon  
\delta\mathbf{S}_{i}  \cdot \mat{\mathcal{J}}^{\mathrm{ani}}{}_{\mkern -22mu ij}\,\,(\epsilon)\cdot\delta\mathbf{S}_{j}
.\label{eq_bilinear3}
\end{align}

The integrands are energy-dependent products of matrices of the same size than the inter-site Green functions and are given by the following forms:

\begin{align}
\mathcal{J}_{ij}^{\mathrm{iso}} (\epsilon) &= A_{ij}(\epsilon)  {t}^{\sigma}_j(\epsilon)A_{ji}(\epsilon)  {t}^{\sigma}_i(\epsilon)
- \left[{\boldsymbol{B}}^\sigma_{ij}(\epsilon)  {t}^{\sigma}_j(\epsilon)\right]\cdot \left[{\boldsymbol{B}}^\sigma_{ji}(\epsilon)  {t}^{\sigma}_i(\epsilon)\right]
\,, \label{eq:integrand_bilinear_Jiso}\\
\boldsymbol{\mathcal{D}}_{ij}(\epsilon)  &= A_{ij}(\epsilon)  {t}^{\sigma}_j(\epsilon){\boldsymbol{B}}^\sigma_{ji}(\epsilon)  {t}^{\sigma}i(\epsilon)
- {\boldsymbol{B}}^\sigma_{ij}(\epsilon)  {t}^{\sigma}_j(\epsilon) A_{ji}(\epsilon)  {t}^{\sigma}_i(\epsilon) \label{eq:integrand_bilinear_DMI}
\,\,\, 
\\
\delta\mathbf{S}_{i}  \cdot \mat{\mathcal{J}}^{\mathrm{ani}}{}_{\mkern -22mu ij}\,\, (\epsilon)\cdot\delta\mathbf{S}_{j} &=
 \left[{\boldsymbol{B}}^\sigma_{ij}(\epsilon) \cdot \delta\mathbf{S}_{j}\right]  {t}^{\sigma}_j(\epsilon)
\left[ {\boldsymbol{B}}^\sigma_{ji}(\epsilon)\cdot \delta\mathbf{S}_{i} \right]{t}^{\sigma}_i(\epsilon) 
+
\left[{\boldsymbol{B}}^\sigma_{ij}(\epsilon) \cdot \delta\mathbf{S}_{i}\right]  {t}^{\sigma}_j(\epsilon)
\left[ {\boldsymbol{B}}^\sigma_{ji}(\epsilon)\cdot \delta\mathbf{S}_{j} \right]{t}^{\sigma}_i(\epsilon) 
\,,
\label{eq:integrand_bilinear_Jani}
\end{align}
where the decomposition of the Green functions introduced in Eq.~\eqref{Green_decomposed} leads to forms different from but still fully equivalent to the usually presented ones~\cite{Udvardi2003,Ebert2009}. Indeed, the elements of the tensor of MEI are  conventionally obtained from Eq.~\ref{eq:energy_2spin} and then combined to calculate the isotropic MEI, DMI and the compass term. 
Interestingly, this  decomposition  permits to identify two contributions to $J$: one is independent of magnetism and SOC while the other one depends on both with at least a quadratic dependence on SOC but being rotationally invariant in spin space owing to the inner product of the different $\mathbf{B}$. $\mathbf{D}$ is, however, linear in $\mathbf{B}$ and consequently in SOC. The direction of the DMI vector is dictated by $\mathbf{B}$ and for lattices with inversion symmetry, I naturally recover that the DMI vanishes because of the symmetry of $A$ and $\mathbf{B}$ with respect to site exchange.

\subsection{Four-spin interaction terms}
The fourth-order term in the Taylor expansion of the energy,  
\begin{align}
\delta E_{4\text{-}\mathrm{spin}} &= - \frac{1}{4\pi} \mathrm{Im}\, \trace \int_{-\infty}^{\epsilon_F} \diff\epsilon  \, \sum_{ijkl} {G}^\mathrm{str}_{ij}(\epsilon) \delta {t}_j(\epsilon){G}^\mathrm{str}_{ji}(\epsilon) \delta {t}_k(\epsilon)
{G}^\mathrm{str}_{kl}(\epsilon) \delta {t}_l(\epsilon){G}^\mathrm{str}_{li}(\epsilon) \delta {t}_i(\epsilon)\, ,
\label{eq:E-4spin}
\end{align}
involves plaquettes of four magnetic moments. 
One notes that, in general, the four-spin contracted-site interaction terms can be recovered from the aforementioned forms by equaling some of the indices, for example by replacing the fourth site $l$ with $j$ to get the three-site terms, while replacing both $(k,l)$ with $(i,j)$ leads to the biquadratic terms. Note that this can lead to vanishing of some of the derived terms and the recovery of bilinear-like terms.

\noindent{\bf Isotropic interaction.} Analogous to the two-spin interaction, $E_{4\text{-}\mathrm{spin}}$ 
gives rise to the conventional isotropic four-spin interactions proportional to $(\mathbf{S}_i\cdot\mathbf{S}_j)(\mathbf{S}_k\cdot\mathbf{S}_l)$ including quadratic and fourth order contributions of the SOC. The isotropic change of the energy reads
\begin{align}
\delta E ^\mathrm{iso}_{ijkl}  &= 
-{\mathcal{B}}_{ijkl} \left[
\left(\delta\mathbf{S}_i \cdot \delta\mathbf{S}_j\right)
\left(\delta\mathbf{S}_k \cdot \delta\mathbf{S}_l\right)
-\left(\delta\mathbf{S}_i \times \delta\mathbf{S}_j\right)
\left(\delta\mathbf{S}_k \times \delta\mathbf{S}_l\right)
\right]
\end{align}
with which one recovers the usual form of the isotropic four-spin energy

\begin{align}
  E ^\mathrm{iso}_{ijkl}  &=
-{\mathcal{B}}_{ijkl} \left[
\left(\mathbf{S}_i \cdot \mathbf{S}_j\right)
\left(\mathbf{S}_k \cdot \mathbf{S}_l\right)
-\left(\mathbf{S}_i \cdot \mathbf{S}_k\right)
\left(\delta\mathbf{S}_j \cdot \mathbf{S}_l\right)
+\left(\mathbf{S}_i \cdot \mathbf{S}_l\right)
\left(\mathbf{S}_j \cdot \mathbf{S}_k\right)
\right]\, ,
\end{align}
with 
\begin{align}
{\mathcal{B}}_{ijkl}  &=  \frac{1}{2\pi}
  \mathrm{Im}\, \trace
\int_{-\infty}^{\epsilon_F} \diff\epsilon  
\left\{A_{ij}(\epsilon)  {t}^{\sigma}_j(\epsilon)A_{jk}(\epsilon)  {t}^{\sigma}_k(\epsilon)A_{kl}(\epsilon)  {t}^{\sigma}_l(\epsilon)A_{li}(\epsilon)  {t}^{\sigma}_i(\epsilon)\right\}\,,
\label{eq:4-spin-iso-interaction}
\end{align}
in zero-order with respect to $\mathbf{B}$. 

\noindent{\bf Four-spin chiral interactions.} The fourth-order term in the Taylor expansion of the energy gives also rise to terms linear in the SOC. The latter are the recently derived~\cite{Brinker2019} and postulated~\cite{Laszloffy2019} four-spin vector-chiral interactions  proportional to  $ (\mathbf{S}_i\cdot\mathbf{S}_j)(\mathbf{S}_k\times\mathbf{S}_l)$ and to the scalar spin-chirality $\mathbf{S}_i \cdot (\mathbf{S}_j\times\mathbf{S}_k)$~\cite{Grytsiuk2020}. Besides these two terms, one additional term proportional to the three-spin vector chirality $\mathbf{S}_i \times (\mathbf{S}_j\times\mathbf{S}_k)$ shows up. I note that the different terms can be rewritten as function of each others. The product of the four Green-function elements in Eq.~\eqref{eq:E-4spin} and the requested linearity in $\mathbf{B}$ leads finally to four terms of the type $\mathbf{B} AAA$, where $\mathbf{B}$ can be placed at four different positions relative to $A$. 
For instance,  $\mathbf{B} AAA$ can be written in various ways depending on how the products involving the Pauli matrices are grouped:
\begin{align}
\delta E^{\mathbf{B}AAA}_{ijkl}  &=- \boldsymbol{\mathcal{C}}_{\bar{i}\bar{j}kl}\cdot[
(\delta\mathbf{S}_{j} \cdot 
\delta\mathbf{S}_{k})
(\delta\mathbf{S}_{l} \times 
\delta\mathbf{S}_{i})
+
(\delta\mathbf{S}_{j} \times 
\delta\mathbf{S}_{k})
(\delta\mathbf{S}_{l} \cdot 
\delta\mathbf{S}_{i})
+
(\delta\mathbf{S}_{j} \times 
\delta\mathbf{S}_{k}) \times
(\delta\mathbf{S}_{l} \times 
\delta\mathbf{S}_{i})
]\label{eq:4SC1}
\end{align}
or
\begin{align} 
\delta E^{\mathbf{B}AAA}_{ijkl}  &=
- (\boldsymbol{\mathcal{C}}_{\bar{i}\bar{j}kl} \cdot  \delta\mathbf{S}_{j}) \left[ 
(\delta\mathbf{S}_{k} \times 
\delta\mathbf{S}_{l}) \cdot \delta\mathbf{S}_{i}
\right] 
- (\boldsymbol{\mathcal{C}}_{\bar{i}\bar{j}kl} \times  \delta\mathbf{S}_{j}) \left[ 
(\delta\mathbf{S}_{k} \cdot 
\delta\mathbf{S}_{l}) \cdot \delta\mathbf{S}_{i}
\right] \\\nonumber
&+ (\boldsymbol{\mathcal{C}}_{\bar{i}\bar{j}kl} \times  \delta\mathbf{S}_{j}) \cdot \left[ 
(\delta\mathbf{S}_{k} \times 
\delta\mathbf{S}_{l}) \times \delta\mathbf{S}_{i}
\right] \,,\label{eq:4SC2}
\end{align}
with 
\begin{align}
\boldsymbol{\mathcal{C}}_{\bar{i}\bar{j}kl} &=
 \frac{1}{2\pi} \mathrm{Re}\,\trace \int_{-\infty}^{\epsilon_F}\!\!\!\! \diff\epsilon \,  
\mathbf{B}_{ij}(\epsilon)  t^{\sigma}_j(\epsilon)A_{jk}(\epsilon)  t^{\sigma}_k(\epsilon)
A_{kl}(\epsilon) t^{\sigma}_l(\epsilon)A_{li}(\epsilon)  t^{\sigma}_i(\epsilon) ,
\end{align}
where the bars on the indices indicate the sites connected by $\mathbf{B}$. It is interesting to note that the integrand showing up in the previous equations can be interpreted as products involving  the bilinear terms $\mathcal{J}^\mathrm{iso}$ and $\boldsymbol{\mathcal{D}}$ (see Eqs.\ref{eq:integrand_bilinear_Jiso}--\ref{eq:integrand_bilinear_Jani}).

In Eq.~\ref{eq:4SC2},  one can identify the three-spin scalar chirality, $\chi_{ijk}=\mathbf{S}_i \cdot (\mathbf{S}_j \times \mathbf{S}_k)$, and if I introduce what I call a three-spin vector chirality $\boldsymbol{\aleph}_{ijk}=\mathbf{S}_i \times (\mathbf{S}_j \times \mathbf{S}_k)$, the four-spin chiral energy linear in $\mathbf{B}$ reads 

\begin{eqnarray}
  E^{\mathbf{B}}_{ijkl}  &=&
  -\boldsymbol{\mathcal{C}}_{\bar{i}\bar{j}kl} \cdot \left[ 
  \mathbf{S}_{j}\, \chi_{ikl} 
  - (\mathbf{S}_{i} \times \mathbf{S}_{j})
 (\mathbf{S}_{k} \cdot \mathbf{S}_{l}) +
 \mathbf{S}_{j}\times
\boldsymbol{\aleph}_{ikl}
  \right]\nonumber\\
 &&-\boldsymbol{\mathcal{C}}_{i\bar{j}\bar{k}l} \cdot \left[
    \mathbf{S}_{k}\, \chi_{jli}
- (\mathbf{S}_{j} \times \mathbf{S}_{k}) ]
(\mathbf{S}_{l} \cdot 
\mathbf{S}_{i}) 
+\mathbf{S}_{k} \times 
 \boldsymbol{\aleph}_{jli} \right] \nonumber \\
&&
- \boldsymbol{\mathcal{C}}_{ij\bar{k}\bar{l}} \cdot \left[
  \mathbf{S}_{l}\,
\chi_{kij}
-   (\mathbf{S}_{k} \times \mathbf{S}_{l})
(\mathbf{S}_{i} \cdot 
 \mathbf{S}_{j}) 
+   \mathbf{S}_{l} \times \boldsymbol{\aleph}_{kij} \right] \nonumber \\
    &&
- \boldsymbol{\mathcal{C}}_{\bar{i}jk\bar{l}} \cdot \left[   \mathbf{S}_{i}\,
\chi_{ljk}
-  (\mathbf{S}_{l} \times \mathbf{S}_{i} )
(\mathbf{S}_{j} \cdot 
 \mathbf{S}_{k}) 
+ \mathbf{S}_{i} \times \boldsymbol{\aleph}_{ljk} \right]\,.
\label{eq:4spin-SCI}
\end{eqnarray}

This equation indicates that the ground state favored by such a four-spin chiral energy is the result of a complex competition of various terms: the three-spin scalar chirality, the three-spin vector chirality and a mix of cross and dot products. For instance, the three-spin scalar chirality pertaining to an equilateral triangular plaquette reaches its largest value for a polar angle $\cos(\theta) = {1}/{\sqrt{3}}$ while the largest length of the three-spin vector chirality is found for for $\cos(\theta) = {\sqrt{7}}/{3}$. By focusing on the first line of 
Eq.~\ref{eq:4spin-SCI}, one notices that the inner product  $\boldsymbol{\mathcal{C}}_{\bar{i}\bar{j}kl} \cdot \boldsymbol{S}_j$ favors a parallel alignement between the four-spin chiral interaction vector and the  magnetic moment $j$. The cross products $\mathbf{S}_{i} \times \mathbf{S}_{j}$, however, drives the spin-configuration towards a state where the two perpendicular magnetic moments $\mathbf{S}_{j} $ and $\mathbf{S}_{i}$ live in a plane perpendicular to  $\boldsymbol{\mathcal{C}}_{\bar{i}\bar{j}kl}$. The third term favors the chiral vector and the three magnetic moments at sites $j,k,l$ to lie in the same plane perpendicular to $\mathbf{S}_{i}$ with $\mathbf{S}_{k}$ and $\mathbf{S}_{l}$ perpendicular to each other while $\boldsymbol{\mathcal{C}}_{\bar{i}\bar{j}kl}$ and $\mathbf{S}_{i}$ would be parallel.  Although, I present the four-spin energy as a sum of three types of terms, as shown in Ref.~\cite{Brinker2020} one can rewrite all of them as function of  the latter one.

While $E^\mathbf{B}$ contains the first-order contributions in terms of $\mathbf{B}$ to $E_{4\text{-}\mathrm{spin}}$, second-order contributions give rise to terms  correcting the dot-product and cross-product of the magnetic moments:
\begin{eqnarray}
E^{\mathbf{B}\mathbf{B}}_{ijkl} &=& E^{\mathbf{B}\mathbf{B}AA}_{ijkl} +
E^{A\mathbf{B}\mathbf{B}A}_{ijkl} +
E^{AA\mathbf{B}\mathbf{B}}_{ijkl} +
E^{A\mathbf{B}A\mathbf{B}}_{ijkl} +
E^{\mathbf{B}A\mathbf{B}A}_{ijkl} +
E^{\mathbf{B}AA\mathbf{B}}_{ijkl}
\end{eqnarray}

As it can be noticed in the following example, where I provide the contribution arising from $\mathbf{B}\mathbf{B} AA$, one finds  products of bilinear terms similar to  $\mathcal{J}^\mathrm{ani}$ multiplying $J^\mathrm{iso}$ or  $\boldsymbol{\mathcal{D}}$ as defined in  Eqs.\ref{eq:integrand_bilinear_Jiso}--\ref{eq:integrand_bilinear_Jani} :
\begin{eqnarray}
E^{\mathbf{B}\mathbf{B}AA}_{ijkl}  &=&  -  \frac{1}{2\pi} \mathrm{Im}\,\trace \int_{-\infty}^{\epsilon_F}\!\!\!\! \diff\epsilon 
  \{ 
  \left( \mathbf{B}_{ij}t_j^{\sigma}  \cdot \mathbf{S}_{j}\right)
\left( \mathbf{B}_{jk}t_k^{\sigma}  \cdot    
  \mathbf{S}_{k}\right)
\left( \mathbf{S}_{l} \cdot      
  \mathbf{S}_{i}\right)
\nonumber\\
&&
-
 \left[\left( \mathbf{B}_{ij}t_j^{\sigma}  \times \mathbf{S}_{j}\right)\cdot
 \left( \mathbf{S}_{l} \times      
  \mathbf{S}_{i}\right)\right]
\left( \mathbf{B}_{jk}t_k^{\sigma}  \cdot    
  \mathbf{S}_{k}\right)
- 
 \left( \mathbf{B}_{ij}t_j^{\sigma}  \cdot \mathbf{S}_{j}\right)
 \left[
 \left( \mathbf{B}_{jk}t_k^{\sigma}  \times    
  \mathbf{S}_{k}\right) \cdot
 \left( \mathbf{S}_{l} \times      
  \mathbf{S}_{i}\right)\right]
  \nonumber\\
&&+ 
 \left[
 \left( \mathbf{B}_{ij}t_j^{\sigma}  \times \mathbf{S}_{j}\right) \times
 \left( \mathbf{B}_{jk}t_k^{\sigma}  \times \mathbf{S}_{k}\right) 
 \right] \cdot
 \left( \mathbf{S}_{l} \times      
  \mathbf{S}_{i}\right)
\} A_{kl}t_l^{\sigma} A_{li}t_i^{\sigma} 
 \,\label{eq:anis-four-spin}.
\end{eqnarray}

The first r.h.s. term looks like the anisotropic bilinear term (Eqs.\ref{eq:integrand_bilinear_Jani}) weighted by a dot product between two magnetic moments of the four-spin plaquette. The second and the third terms can be rewritten in a form similar to the first one:

\begin{align}
  \left[\left( \mathbf{B}_{ij}t_j^{\sigma}  \times \mathbf{S}_{j}\right)\cdot
 \left( \mathbf{S}_{l} \times      
  \mathbf{S}_{i}\right)\right]
\left( \mathbf{B}_{jk}t_k^{\sigma}  \cdot    
  \mathbf{S}_{k}\right) &=
  \left[\left( \mathbf{B}_{ij}t_j^{\sigma}  \cdot \mathbf{S}_{l}\right)
 \left( \mathbf{S}_{j} \cdot      
  \mathbf{S}_{i}\right)
  -
  \left( \mathbf{B}_{ij}t_j^{\sigma}  \cdot \mathbf{S}_{i}\right)
 \left( \mathbf{S}_{j} \cdot      
  \mathbf{S}_{l}\right)
  \right]
\left( \mathbf{B}_{jk}t_k^{\sigma}  \cdot    
  \mathbf{S}_{k}\right)
\end{align}
and
\begin{align}
 \left( \mathbf{B}_{ij}t_j^{\sigma}  \cdot \mathbf{S}_{j}\right)
 \left[
 \left( \mathbf{B}_{jk}t_k^{\sigma}  \times    
  \mathbf{S}_{k}\right) \cdot
 \left( \mathbf{S}_{l} \times      
  \mathbf{S}_{i}\right)\right] &=
   \left( \mathbf{B}_{ij}t_j^{\sigma}  \cdot \mathbf{S}_{j}\right)
 \left[
 \left( \mathbf{B}_{jk}t_k^{\sigma}  \cdot    
  \mathbf{S}_{l}\right) 
 \left( \mathbf{S}_{k} \cdot      
  \mathbf{S}_{i}\right)
  -
  \left( \mathbf{B}_{jk}t_k^{\sigma}  \cdot    
  \mathbf{S}_{i}\right) 
 \left( \mathbf{S}_{k} \cdot     
  \mathbf{S}_{l}\right)
  \right] 
\end{align}
while the fourth term can be written as 
\begin{align}
   \left[
 \left( \mathbf{B}_{ij}t_j^{\sigma}  \times \mathbf{S}_{j}\right) \times
 \left( \mathbf{B}_{jk}t_k^{\sigma}  \times \mathbf{S}_{k}\right) 
 \right] \cdot
 \left( \mathbf{S}_{l} \times      
  \mathbf{S}_{i}\right)
  &= [\mathbf{B}_{ij}t_j^{\sigma} \cdot \left( \mathbf{S}_{j} \times \mathbf{S}_{l} \right)  ] 
  [\mathbf{B}_{jk}t_k^{\sigma} \cdot \left( \mathbf{S}_{k}  \times \mathbf{S}_{i} \right)   ] \nonumber\\
  &-
   [\mathbf{B}_{ij}t_j^{\sigma} \cdot  \left(\mathbf{S}_{j}  \times \mathbf{S}_{i} \right)  ] 
  [\mathbf{B}_{jk}t_k^{\sigma} \cdot 
  \left( \mathbf{S}_{k}  \times \mathbf{S}_{l} \right)  ]\,.
\end{align}

\subsection{Six-spin interaction term}
The sixth order term obtained from  Eq.~\eqref{Lloyd} involves a plaquette of six magnetic moments. One expects the largest contribution to be the isotropic, rotationally invariant one, as this term can be finite without spin-orbit interaction. Although being isotropic, one finds contributions involving products of three-spin scalar chiralities, or of three-spin vector chiralities:
\begin{align}
\delta E_{ijklmn}^\text{iso} &=
-
\varkappa^{6\text{-}\mathrm{spin}}_{ijklmn} \left[
\left( \delta\mathbf{S}_i\cdot \delta\mathbf{S}_j
\right)
\left( \delta\mathbf{S}_k\cdot \delta\mathbf{S}_l
\right)
\left( \delta\mathbf{S}_m\cdot \delta\mathbf{S}_n
\right)
- \left( \delta\mathbf{S}_i\cdot \delta\mathbf{S}_j
\right)
\left( \delta\mathbf{S}_k\times \delta\mathbf{S}_l
\right)\cdot
\left( \delta\mathbf{S}_m\times \delta\mathbf{S}_n
\right) \right. \nonumber\\
&\left.
\,\,\,\,\,\,\,\,\,\,\,\,\,\,\,\,\,\,- \left( \delta\mathbf{S}_i\times \delta\mathbf{S}_j
\right)
\left( \delta\mathbf{S}_k\cdot \delta\mathbf{S}_l
\right)\cdot
\left( \delta\mathbf{S}_m\times \delta\mathbf{S}_n
\right)
- \left( \delta\mathbf{S}_i\times \delta\mathbf{S}_j
\right)\cdot
\left( \delta\mathbf{S}_k\times \delta\mathbf{S}_l
\right)\cdot
\left( \delta\mathbf{S}_m\cdot \delta\mathbf{S}_n
\right)\right] \,,
\end{align}
and thus
\begin{align}
E_{ijklmn}^\text{iso} &=
-
\varkappa^{6\text{-}\mathrm{spin}}_{ijklmn} [
\left( \mathbf{S}_i\cdot \mathbf{S}_j
\right)
\left( \mathbf{S}_k\cdot\mathbf{S}_l
\right)
\left( \mathbf{S}_m\cdot \mathbf{S}_n
\right)
\nonumber\\
&
- 
 \chi_{imn}\chi_{jkl} - \chi_{kmn}\chi_{lij} - \chi_{mkl}\chi_{nij} 
 \nonumber\\
&
 - \boldsymbol{\aleph}_{ikl}\cdot\boldsymbol{\aleph}_{jmn}
- \boldsymbol{\aleph}_{kij}\cdot\boldsymbol{\aleph}_{lmn}
- \boldsymbol{\aleph}_{mij}\cdot\boldsymbol{\aleph}_{nkl}]
\,,
\label{eq:E_CC_ijkijk}
\end{align}
with $\chi_{ijk}=\mathbf{S}_i \cdot (\mathbf{S}_j \times \mathbf{S}_k)$ and $\boldsymbol{\aleph}_{ijk}=\mathbf{S}_i \times (\mathbf{S}_j \times \mathbf{S}_k)$. 
In general, the effective 6-spin interaction is given by
 \begin{align}
\varkappa^{6\text{-}\mathrm{spin}}_{ijklmn} &= \frac{1}{3\pi} \mathrm{Im}\, \trace
\int_{-\infty}^{\epsilon_F} \diff\epsilon \,  {A}_{ij}(\epsilon)  t^{\sigma}_j(\epsilon){A}_{jk}(\epsilon)  t^{\sigma}_k(\epsilon)
{A}_{kl}(\epsilon) t^{\sigma}_l(\epsilon)
{A}_{lm}(\epsilon)  t^{\sigma}_m(\epsilon){A}_{mn}(\epsilon)  t^{\sigma}_n(\epsilon)
{A}_{ni}(\epsilon)  t^{\sigma}_i(\epsilon)\, . 
\end{align}

\section{Rashba model}
Here I discuss the derived   multi-spin interactions considering  localized spins embedded in 1D and 2D electron gas, with spin-orbit coupling added in the form of Rashba coupling~\cite{Rashba1960,Bychkov1984}. The conduction electrons are then either confined along a given direction (1D) or in the $xy$-plane (2D). The effective electric field emerging from spin-orbit coupling points along the $z$-axis. The Hamiltonian of both systems consists of the kinetic 
energy of the free electrons to which  a linear term in momentum $\vec{p}$ is added:
\begin{align}
\boldsymbol{H} &= \frac{ \hbar^2 }{2m^*} \boldsymbol{\nabla}^2\,\boldsymbol{\sigma}_0- 
\alpha( -\iu {\hbar}
\boldsymbol{\nabla} \times \bar{z})\cdot \boldsymbol{\sigma}\,,
\label{rashba_hamiltonian}
\end{align}

where $\alpha$ is the so-called Rashba parameter representing  
the strength of the spin-orbit  interaction, $\hat{z}$ is the unit vector along the $z$-axis, $m^*$ is the effective mass of the electron,  $\sigma_0$ is the identity matrix and $\boldsymbol{\sigma}$ is the vector of Paul spin matrices. 

Similarly to Ref.~\cite{Imamura2004}, the potential of each of  the localized spins is described via the  $s-d$ interaction $J^{sd}$, which has units of energy multiplying length to the power of dimensionality, such that the change of the magnetic part of the potential upon rotation of the spin moment is given by
\begin{align}
  \delta V (\boldsymbol{r}) &= \sum_i J^{sd}_i  \delta({\boldsymbol{r}-\boldsymbol{r}_i})\,\mathbf{S}_i\cdot\sigma\,.
\end{align}

In the following, I use the $s-d$ potential instead of the energy-dependent $t$-matrix in order to get analytical forms of the multi-spin interactions similarly to what is done in the RKKY approximation. This means that I will use the expansion given by Eq.~\ref{Lloyd1} based on the full Green function instead of Eq.~\ref{Lloyd}, where the structural Green function is utilized.

\subsection{Rashba model -- One dimension}
The  Rashba Green function corresponding to the previous Hamiltonian for the 1DEG reads
\begin{align}
G(\boldsymbol{r}_{ij};\varepsilon + \iu 0^+ )& = 
A_{ij} (\varepsilon + \iu 0^+) + B_{ij}(\varepsilon + \iu 0^+)\, \hat{\beta}_{ij}\cdot \boldsymbol{\sigma}\,,
\end{align}

where $\boldsymbol{r}_{ij} = r_{ij}(\cos\beta_{ij},\sin\beta_{ij})$ is the vector connecting sites $i$ and $j$, while $\hat{\beta}_{ij} = (\sin\beta_{ij},-\cos\beta_{ij})$ is the unit vector perpendicular to $\boldsymbol{r}_{ij}$. Defining $q = \sqrt{\frac{2m}{\hbar^2}\varepsilon + k_R^2}$ with $k_R = \frac{m^*}{\hbar^2}\alpha$, the components of the Green functions are~\cite{Imamura2004}:
\begin{align}
A_{ij} (\varepsilon + \iu 0^+) &= -\iu \frac{m^*}{\hbar^2} \frac{e^{\iu q |r_{ij}|}}{q+ \iu 0^+} \cos{(k_R r_{ij})}\,\,\, \text{and}\,\,\,B_{ij} (\varepsilon + \iu 0^+) =  \frac{m^*}{\hbar^2} \frac{e^{\iu q |r_{ij}|}}{q+ \iu 0^+} \sin{(k_R r_{ij})}\,
\end{align}

Now, I can proceed with the multi-spin interactions  by plugging in $A$ and $B$ into the MEI derived in Section II. I  start with the basic bilinear types~\cite{Imamura2004,Bouaziz2017}.

\subsubsection{Two-spin interactions}
\hfill \\
Within the 1DEG  Rashba model, utilizing Eqs.~\ref{eq_bilinear1}--\ref{eq:integrand_bilinear_Jani}, I recover the bilinear magnetic interactions that were already derived in Ref.~\cite{Imamura2004}:
\begin{align}
J_{ij}^{\mathrm{iso}} &= \mathcal{F}^{(1\text{D})}_{ij} \cos{(2k_R r_{ij})}\, ,\label{eq:MEI_bilinear_Rashba_1D_a}
\\
\noalign{\noindent \vspace{2\jot}} 
\delta\mathbf{S}_{i}  \cdot \mat{J}^{\mathrm{ani}}{}_{\mkern -22mu ij}\cdot\delta\mathbf{S}_{j} &= \mathcal{F}^{(1\text{D})}_{ij}
(1 - \cos{(2k_R r_{ij})} )
\left[{\hat{\beta}}_{ij} \cdot \delta\mathbf{S}_{j}\right] 
\left[ {\hat{\beta}}_{ij}\cdot \delta\mathbf{S}_{i} \right] 
,\label{eq:MEI_bilinear_Rashba_1D_b}\\
\noalign{\noindent and \vspace{2\jot}} 
\mathbf{D}_{ij} &= \mathcal{F}^{(1\text{D})}_{ij} \sin{(2k_R r_{ij})} \hat{\beta}_{ij}\,.\label{eq:MEI_bilinear_Rashba_1D_c}
\end{align}
Within the RKKY approximation, the direction of the DMI is along ${\hat{\beta}}_{ij}$, i.e. perpendicular to the bond connecting the two sites $i$ and $j$. Naturally, without SOC, $\mathbf{D}$ and $\mat{J}^{\mathrm{ani}}$ vanish. For small $k_R r_{ij}$, the DMI is linear with the Rashba parameter $\alpha$ while the compass-term interaction shows a quadratic dependence similarly to the SOC correction to the isotropic MEI. These dependencies change at large $k_R r_{ij}$, where the various bilinear interactions can be of the same order of magnitude with a decay dictated by the range function $\mathcal{F}^{(1\text{D})}_{ij}$, which is in fact the isotropic bilinear MEI before application of the SOC~\cite{Yafet1987,Litvinov1998}:
\begin{align}
\mathcal{F}^{(1\text{D})}_{ij} &=- {J^{sd}_iJ^{sd}_j} \frac{m^{*2}}{\pi\hbar^4}
\mathrm{Im} \, \trace \int_{-\infty}^{\epsilon_F} \diff\epsilon\, 
\frac{e^{\iu 2 q r_{ij}}}{(q+ \iu 0^+)^2} \nonumber\\
&= - {J^{sd}_iJ^{sd}_j} \frac{m^{*}}{\pi\hbar^2}\left[ \text{Si}(2q_F r_{ij}) - \frac{\pi}{2}\right]\,,
\end{align}
where $\text{Si}()$ is the sine integral function. Owing to the asymptotic behavior of the sine function: $Si(x) \simeq  \frac{\pi}{2} - \frac{\cos{(x)}}{x}\left( 1 - \frac{2!}{x^2} + \frac{4!}{x^4} - \frac{6!}{x^6} +... \right)- \frac{\sin{(x)}}{x}\left( \frac{1}{x} - \frac{3!}{x^3} + \frac{5!}{x^5} - \frac{7!}{x^7} +...\right)$, the range function behaves like:
\begin{eqnarray}
\mathcal{F}^{(1\text{D})}_{ij} &\simeq& {J^{sd}_iJ^{sd}_j} \frac{m^{*}}{\pi\hbar^2}\left[ 
 \frac{\cos{(2q_F r_{ij})}}{2q_F r_{ij}}\left( 1 - \frac{1}{2q_F^2 r_{ij}^2}  \right)
 +\frac{\sin{(2q_F r_{ij})}}{4q_F^2 r_{ij}^2}
\right]\,.\label{eq:asymp_bilinear_range_function_1D}
\end{eqnarray}
Thus, the bilinear interactions are characterized by various decays, $r_{ij}^{-1}$ being the lowest one, but interference effects are expected with the term proportional to $r_{ij}^{-2}$. The wave length of the oscillations is expected to be complex depending on the interatomic distances. It is given by $2 \lambda_F = \pi /q_F$ at small $q_F r$ before interference effects kick in, which originate from either or both the 1D nature of the electron gas (Eq.~\ref{eq:asymp_bilinear_range_function_1D}) and from SOC  (Eqs.~\ref{eq:MEI_bilinear_Rashba_1D_a},~\ref{eq:MEI_bilinear_Rashba_1D_b} and~\ref{eq:MEI_bilinear_Rashba_1D_c}).

\subsubsection{Four-spin interactions}
\hfill \\
\noindent{\textbf{The isotropic four-spin interaction.}} As derived in Eq.~\ref{eq:4-spin-iso-interaction}, this interaction involves a product of four cosine functions that depend on $k_R$:
\begin{eqnarray}
{\mathcal{B}}_{ijkl}  
&=& \mathcal{F}^{(1\text{D})}_{ijkl}
\cos{(k_R r_{ij})}\cos{(k_R r_{jk})}\cos{(k_R r_{kl})}\cos{(k_R r_{li})}\,.
\end{eqnarray}
For small $k_R r$, the isotropic four-spin interaction experiences corrections with even power of the Rashba coupling parameter $\alpha$. The lowest dependence being quadratic in this regime, this implies a SOC contribution to the magnetic energy that is similar to the isotropic bilinear contribution . Also, and in analogy to the latter term,  a range function takes care of the distance-dependent behavior of the four-spin interaction:
\begin{align} 
\mathcal{F}^{(1\text{D})}_{ijkl} &=   {J^{sd}_iJ^{sd}_jJ^{sd}_kJ^{sd}_l} \frac{m^{*4}}{2\pi\hbar^8} 
\mathrm{Im}\, \trace
\int_{-\infty}^{\epsilon_F} \diff\epsilon  \frac{e^{\iu q L_{ijkl}}}{(q+ \iu 0^+)^4}\nonumber\\
&= - {J^{sd}_iJ^{sd}_jJ^{sd}_kJ^{sd}_l}\frac{m^{*3}}{2\pi\hbar^6} 
\left[
\frac{1}{2}\frac{\sin{(q_FL_{ijkl})}}{q_F^2}
+\frac{L_{ijkl}}{2}\frac{\cos{(q_FL_{ijkl})}}{q_F}
+\frac{L_{ijkl}^2}{2}\left(\text{Si}(q_F L_{ijkl}) -\frac{\pi}{2} \right)
\right]\,.
\end{align}
Here I define $L_{ijkl} = r_{ij}+r_{jk}+r_{kl}+r_{li}$ as the perimeter of the 1D plaquette $(ijkl)$. Using once more the asymptotic form of the Sine integral function, the long-range behavior of  $\mathcal{F}^{(1D)}_{ijkl}$ simplifies to:
\begin{align}
 \mathcal{F}^{(1\text{D})}_{ijkl} &\simeq - {J^{sd}_iJ^{sd}_jJ^{sd}_kJ^{sd}_l}\frac{m^{*3}}{2\pi\hbar^6} 
\left[\frac{\cos(q_F L_{ijkl})}{q_F^3L_{ijkl}} +
3\frac{\sin(q_F L_{ijkl})}{q_F^4 L_{ijkl}^2} 
\right]\,,
\end{align}
which indicates that the decay of the interaction is similar to the bilinear ones and is related to the distance spanned by the scattered electrons involved in the processes giving rise to the MEI. Depending on $q_F$ and on the magnitude of the $s-d$ interaction, the isotropic four-spin interaction can be of the same order of magnitude than the bilinear MEI. Compared to the latter, the wavelength of the oscillations decreases for the four-spin interaction. Assuming magnetic moments equidistant by $R$, the range function becomes:
\begin{align}
 \mathcal{F}^{(1\text{D})}_{ijkl} &\simeq
 - {J^{sd}_iJ^{sd}_jJ^{sd}_kJ^{sd}_l}\frac{m^{*3}}{2\pi\hbar^6} 
\left[
\frac{1}{2}\frac{\sin{(4 q_F R)}}{q_F^2}
+2{R}\frac{\cos{(4q_F R)}}{q_F}
+8{R^2}\left(\text{Si}(4 q_F R) -\frac{\pi}{2} \right)
\right]\nonumber \\
 &
 - {J^{sd}_iJ^{sd}_jJ^{sd}_kJ^{sd}_l}\frac{m^{*3}}{2\pi\hbar^6} 
\left[\frac{\cos(4 q_F R)}{4q_F^3 R} +
3\frac{\sin(4 q_F R)}{16 q_F^4 R^2} 
\right]\,,
\end{align}

\noindent{\textbf{The four-spin chiral interaction.}} The chiral interaction that is linear in $\mathbf{B}$ has the following form
\begin{align}
\boldsymbol{\mathcal{C}}_{\bar{i}\bar{j}kl} &= - \mathcal{F}^{(1\text{D})}_{ijkl} \sin{(k_R r_{ij})}\cos{(k_R r_{jk})}\cos{(k_R r_{kl})}\cos{(k_R r_{li})} \hat{\beta}_{ij}\,,
\end{align}
whose direction is perpendicular to the bond connecting the sites mediating the Green function $\mathbf{B}$ similarly to the DMI vector. Note that this is particular to the RKKY approximation based on the Rashba model since in general, the direction of $\mathbf{\mathcal{C}}_{\bar{i}\bar{j}kl}$ can be more complex as demonstrated in Ref.~\cite{Brinker2020}.  
In comparison to the isotropic four-spin interaction, one of the cosines is replaced by a sine function. For small $k_R r$, similarly to the DMI a linear dependence with respect to the Rashba coupling parameter $\alpha$ is expected for relatively short distances. Therefore, increasing SOC  would permit to increase the amplitude of such chiral four-spin interaction. Since it is quadratic in $\mathbf{B}$, the anisotropic chiral four-spin interaction involves  two sine functions depending on $\alpha$ and therefore the short-distance behavior is quadratic with respect to SOC:
 \begin{eqnarray}
  \mathcal{C}^\text{ani}_{\bar{i}\bar{j}\bar{k}l} 
&=& - \mathcal{F}_{ijkl}^{(1\text{D})} \, \sin{(k_R r_{ij})}
 \, \sin{(k_R r_{jk})} \, \cos{(k_R r_{kl})}\, \cos{(k_R r_{li})}\,.
 \end{eqnarray}
This parametrizes the energy contributions such as the one given in Eq.~\ref{eq:anis-four-spin}, which in the Rashba model reads:

\begin{align}
E^{\mathbf{B}\mathbf{B}AA}_{ijkl} &= - \mathcal{C}^\text{ani}_{\bar{i}\bar{j}\bar{k}l}  
  \left\{
  \left( \hat{\beta}_{ij}t_j^s  \cdot \mathbf{S}_{j}\right)
\left( \hat{\beta}_{jk}t_k^s  \cdot    
  \mathbf{S}_{k}\right)
\left( \mathbf{S}_{l} \cdot      
  \mathbf{S}_{i}\right) \right.
\nonumber\\&
-
 \left(\left( \hat{\beta}_{ij}t_j^s  \times \mathbf{S}_{j}\right)\cdot
 \left( \mathbf{S}_{l} \times      
  \mathbf{S}_{i}\right)\right)
\left( \hat{\beta}_{jk}t_k^s  \cdot    
  \mathbf{S}_{k}\right)
- 
 \left( \hat{\beta}_{ij}t_j^s  \cdot \mathbf{S}_{j}\right)
 \left(
 \left( \hat{\beta}_{jk}t_k^s  \times    
  \mathbf{S}_{k}\right) \cdot
 \left( \mathbf{S}_{l} \times      
  \mathbf{S}_{i}\right)\right)
  \nonumber \\
  &
+ 
\left. \left(
 \left( \hat{\beta}_{ij}t_j^s  \times \mathbf{S}_{j}\right) \times
 \left( \hat{\beta}_{jk}t_k^s  \times \mathbf{S}_{k}\right) 
 \right) \cdot
 \left( \mathbf{S}_{l} \times      
  \mathbf{S}_{i}\right)
\right\} A_{kl}t_l^s A_{li}t_i^s  \,.
\end{align}

\subsubsection{{Six-spin interactions}} 
\hfill \\
The isotropic six-spin interaction is given by 
\begin{align}
\varkappa^\text{6\text{-spin}}_{ijklmn} &=  \mathcal{F}_{ijklmn}^{(1\text{D})}  \cos{(k_R r_{ij})}\,\cos{(k_R r_{jk})}\,\cos{(k_R r_{kl})}\,\cos{(k_R r_{lm})}\,\cos{(k_R r_{mn})}\,\cos{(k_R r_{ni})} \, , 
\end{align}
where 
\begin{align}
\mathcal{F}_{ijklmn}^{(1\text{D})} &= 
{J^{sd}_i J^{sd}_j J^{sd}_k J^{sd}_l J^{sd}_m J^{sd}_n }\frac{ {m^*}^5}{3\pi\hbar^{10}}
 \left[
 \frac{1}{4}\frac{\sin(q_F L_{ijklmn})}{q_F^4} +\frac{L_{ijklmn}}{12} \frac{\cos{(q_F L_{ijklmn})}}{q_F^3}
 \right.
 \nonumber\\
 &
 \left.
 -\frac{L_{ijklmn}^2}{12}\left(
\frac{1}{2}\frac{\sin{(q_FL_{ijklmn})}}{q_F^2}
+\frac{L_{ijklmn}}{2}\frac{\cos{(q_FL_{ijklmn})}}{q_F}
+\frac{L_{ijklmn}^2}{2}\left(\text{Si}(q_F L_{ijklmn}) -\frac{\pi}{2} \right)
\right)
 \right]\,,
\end{align}
which in the asymptotic long-range regime simplifies to
\begin{align}
  \mathcal{F}^{(1\text{D})}_{ijklmn} &\simeq
  {J^{sd}_i J^{sd}_j J^{sd}_k J^{sd}_l J^{sd}_m J^{sd}_n }\frac{ {m^*}^5}{3\pi\hbar^{10}} 
  \left(\frac{\cos(q_F L_{ijklmn})}{q_F^5 L_{ijklmn}} 
  +\frac{5\sin(q_F L_{ijklmn})}{q_F^6 L_{ijklmn}^2}
  \right).
\end{align}
For completeness, I give the form of the range function for equidistant atoms:
\begin{align}
  \mathcal{F}^{(1\text{D})}_{ijklmn} &\simeq
  {J^{sd}_i J^{sd}_j J^{sd}_k J^{sd}_l J^{sd}_m J^{sd}_n }\frac{ {m^*}^5}{3\pi\hbar^{10}} 
  \left(\frac{\cos(6 q_F R)}{6 q_F^5 R} 
  +\frac{5\sin(6 q_F R)}{36 q_F^6 R^2}
  \right).
\end{align}

Interestingly, the decay function is similar to that of the bilinear and four-spin interactions highlighting their potential relevance at large distances.

To summarize the 1D case,  the multi-spin interactions similarly decay like $L^{-1}$. The modification of $q_F$ can considerably affect the amplitude of the interactions owing to their  $q_F^{N-1}$ dependence, $N$ being the number of interacting magnetic moments defining the   investigated plaquette. Moreover, the chiral multi-spin interactions are expected to be at short distances linear, at least, with the Rashba SOC.

\subsection{Rashba model -- Two dimension}
In the 2D case,  $A$ and $B$ defining the Green function are given by linear 
combinations of Hankel functions of zero and first order, respectively: 
\begin{align}
A_{ij}(\varepsilon+\iu 0^+) &= -\frac{\iu m^*}{4\hbar^2}\left[ \left(1+\frac{k_R}{q}\right)\,H_{0}\left[(q+k_R)r_{ij}+\iu 0^+\right] 
+\left(1-\frac{k_R}{q}\right)\,H_0\left[(q-k_R)r_{ij} +\iu 0^+\right]\right]
\nonumber\\
B_{ij}(\varepsilon+\iu 0^+) &= +\frac{ m^*}{4\hbar^2}\left[ \left(1+\frac{k_R}{q}\right)\,H_1\left[(q+k_R)r_{ij}+\iu 0^+\right] 
-\left(1-\frac{k_R}{q}\right)\,H_1\left[(q-k_R)r_{ij} +\iu 0^+\right]\right]\,.
\label{g01_2D}
\end{align}

Similarly to what was done for the 1D case, I proceed to the dervation of the multi-spin MEI in the RKKY approximation. The functional dependence with respect to the magnetic moments is not affected and all the physics is primarily encoded within the range functions, $\mathcal{F}_{ijkl}^{\text(2D)}$, i.e. the MEI without the Rashba SOC, which are listed below in the asymptotic regimes, i.e. when $qr \gg 1$ and $k_R \ll q$. There, the Hankel functions are 
$H_0(x)\simeq \sqrt{\frac{2}{\pi x}}
\,e^{\iu(x-\frac{\pi}{4})}$ and $H_1(x)\simeq \sqrt{\frac{2}{\pi x}}\, 
e^{\iu(x-\frac{3\pi}{4})}$ for large $x$,
which allow to write 
\begin{align}
A_{ij}(\varepsilon+\iu 0^+) &\simeq  
 -\frac{ \iu m^*}{\hbar^2}\frac{1}{\sqrt{2\pi q r_{ij}}} e^{\iu(q r_{ij}-\frac{\pi}{4})} 
 \cos{k_R r_{ij}} \,\,\, \text{and}\,\,\,
 B_{ij}(\varepsilon+\iu 0^+) \simeq
 \frac{ 
 m^*}{\hbar^2}\frac{1}{\sqrt{2\pi q r_{ij}}} e^{\iu(q r_{ij}-\frac{\pi}{4})} 
 \sin{k_R r_{ij}}
\quad ,
\label{g01_1D}
\end{align}

\noindent{\textbf{Two-spin interactions.}}
Here the two-spin range function 
\begin{eqnarray}
\mathcal{F}_{ij}^\text{(2D)} &\simeq&  J_i^{sd} J_j^{sd}
\frac{m^*}{4\pi^2\hbar^2}\,
\frac{\sin(2q_F r_{ij})}{r_{ij}^2} \,
\end{eqnarray}
gives the usual decay function proportional to $r_{ij}^2$ for the 2D electron gas.

\noindent{\textbf{Four-spin interactions.}}
In the asymptotic regime the four-spin range function is similar to the one characterizing the   two-spin 1D interactions
\begin{align}
    \mathcal{F}_{ijkl}^\text{(2D)}  &\simeq -
 J_i^{sd}J_j^{sd}J_k^{sd}J_l^{sd}
 \frac{ {m^*}^3}{8\pi^3\hbar^6} 
 \frac{1}{\sqrt{P_{ijkl}}}
\left[\text{Si}(q_F L_{ijkl}) - \frac{\pi}{2}\right] \nonumber\\
&\simeq
J_i^{sd}J_j^{sd}J_k^{sd}J_l^{sd}
 \frac{ {m^*}^3}{8\pi^3\hbar^6} 
 \frac{1}{\sqrt{P_{ijkl}}}
\left[ 
 \frac{\cos{(q_F L_{ijkl})}}{q_F L_{ijkl}}\left( 1 - \frac{1}{q_F^2 L_{ijkl}^2}  \right)
 +\frac{\sin{(q_F L_{ijkl})}}{q_F^2 L_{ijkl}^2}
\right],
\end{align}
with the already defined perimeter $L_{ijkl} = r_{ij}+r_{jk}+r_{kl}+r_{li}$ and the product of interatomic distances $P_{ijkl} = r_{ij}r_{jk}r_{kl}r_{li}$. The slowest decay function of the  four-spin interactions is given by $(\sqrt{P}L)^{-1}$, which simiplifies to $R^{-3}$ when assuming a plaquette of equidistant  atoms ($r_{ij} = r_{jk} = r_{kl} = r_{li} = R$ ), in which case the range function becomes
\begin{align}
    \mathcal{F}_{ijkl}  &\simeq 
    J_i^{sd}J_j^{sd}J_k^{sd}J_l^{sd}
 \frac{ {m^*}^3}{8\pi^3\hbar^6} 
 \frac{1}{R^2}
  \left[\text{Si}(4 q_F R ) - \frac{\pi}{2}\right]\nonumber\\
  &\simeq -
  J_i^{sd}J_j^{sd}J_k^{sd}J_l^{sd}
 \frac{ {m^*}^3}{8\pi^3\hbar^6} 
 \frac{1}{R^2}
  \left[
   \frac{\cos{(4 q_F R)}}{4q_F R}\left( 1 - \frac{1}{16q_F^2 R^2}  \right)
 +\frac{\sin{(4q_F R)}}{16 q_F^2 R^2}
  \right]
.
\end{align}
Thus and in contrast to the 1D-case, the four-spin interactions are expected to decay faster than the bilinear ones. A similar observation can be made for the following six-spin interactions.

\noindent{\textbf{Six-spin interactions.}} 
Here, the range function is approximately given by 
\begin{align}
\mathcal{F}_{ijklmn}^\text{(2D)} &\simeq   J_i^{sd}J_j^{sd}J_k^{sd}J_l^{sd}J_m^{sd}J_n^{sd} \frac{ {m^*}^5}{24\pi^4\hbar^{10}}\frac{1}{\sqrt{P_{ijklmn}}}  \left( \frac{\cos{(q_FL_{ijklmn})}}{q_F} + L_{ijklmn} \left[\text{Si}(q_FL_{ijklmn}) -\frac{\pi}{2}\right]\right)\nonumber\\
&\simeq
 J_i^{sd}J_j^{sd}J_k^{sd}J_l^{sd}J_m^{sd}J_n^{sd} 
 \frac{ {m^*}^5}{24\pi^4\hbar^{10}}\frac{1}{L_{ijkl}\sqrt{P_{ijklmn}}} \left[ 
 2\frac{\cos(q_F L_{ijkl}) }{q_F^3}
 -\frac{\sin{(q_F L_{ijkl})}}{q_F^2}
\right]
\end{align}
and if the atoms in the plaquette are equidistant as they would be in an hexagonal lattice ($r_{ij} = r_{jk} = r_{kl} = r_{lm} = r_{mn} = r_{ni} =  R$ ), the range function becomes
\begin{align}
\mathcal{F}_{ijklmn} &\simeq J_i^{sd}J_j^{sd}J_k^{sd}J_l^{sd}J_m^{sd}J_n^{sd}  \frac{ {m^*}^5}{24\pi^4\hbar^{10}}\frac{1}{ R^3}  \left( \frac{\cos{(6q_F R)}}{q_F} + 6R \left[\text{Si}(6 q_F R) -\frac{\pi}{2}\right]\right)\nonumber \\
&\simeq J_i^{sd}J_j^{sd}J_k^{sd}J_l^{sd}J_m^{sd}J_n^{sd} 
  \frac{ {m^*}^5}{144\pi^4\hbar^{10}}\frac{1}{R^4}  \left[ 
 2\frac{\cos(6q_F R)}{q_F^3} - \frac{\sin(6q_F R)}{q_F^2}
 \right]
\end{align}

One notices that decay function pertaining to the 2D interactions is given by $(P^{\frac{1}{2}}L)^{-1}$ which is different from the 1D case because of the additional factor $P^{\frac{1}{2}}$ that simplifies to $R^{\frac{N}{2}}$ for $N$ equidistant magnetic moments. This ignites a strong difference in the long-range behavior since the power-law decay associated to the 1D electon gas is independent from the number of atomic spins, while it increases linearly in the 2D case. Similarly to the 1D interactions, the higher-order 2D magnetic interactions are expected to change when modifying  $q_F$ but the power-law dependence is different, $\approx q_F^{N-2}$ for 2D instead of $q_F^{N-1}$.  This points to the possibility of manipulating the impact and magnitude of the magnetic interactions by controlling either the position of the magnetic moments or the Fermi energy.

To summarize, one can state that in general the decay of the multi-spin MEI can be cast into the following formula with which the behavior of the two-spin interactions is recovered ($\approx q_F^{d-2}R^{-d}$):
\begin{eqnarray}
\text{Multi-spin MEI}&\approx& \{q_F^{(N-d)}P^{\frac{1}{2}(d-1)}L\}^{-1}  \approx \{q_F^{(N-d)}R^{\left[1+\frac{N}{2}(d-1)\right]}N\}^{-1} ,
\end{eqnarray}
with $d = 1$ or 2 depending on the dimension of the electron gas mediating the interactions. Noteworthy is the impact of SOC on amplitude of the chiral MEI. It is expected to be at least linear with $\alpha$ independently from the nature of the electron gas.

\begin{figure*}
\centering
\includegraphics[width=\columnwidth]{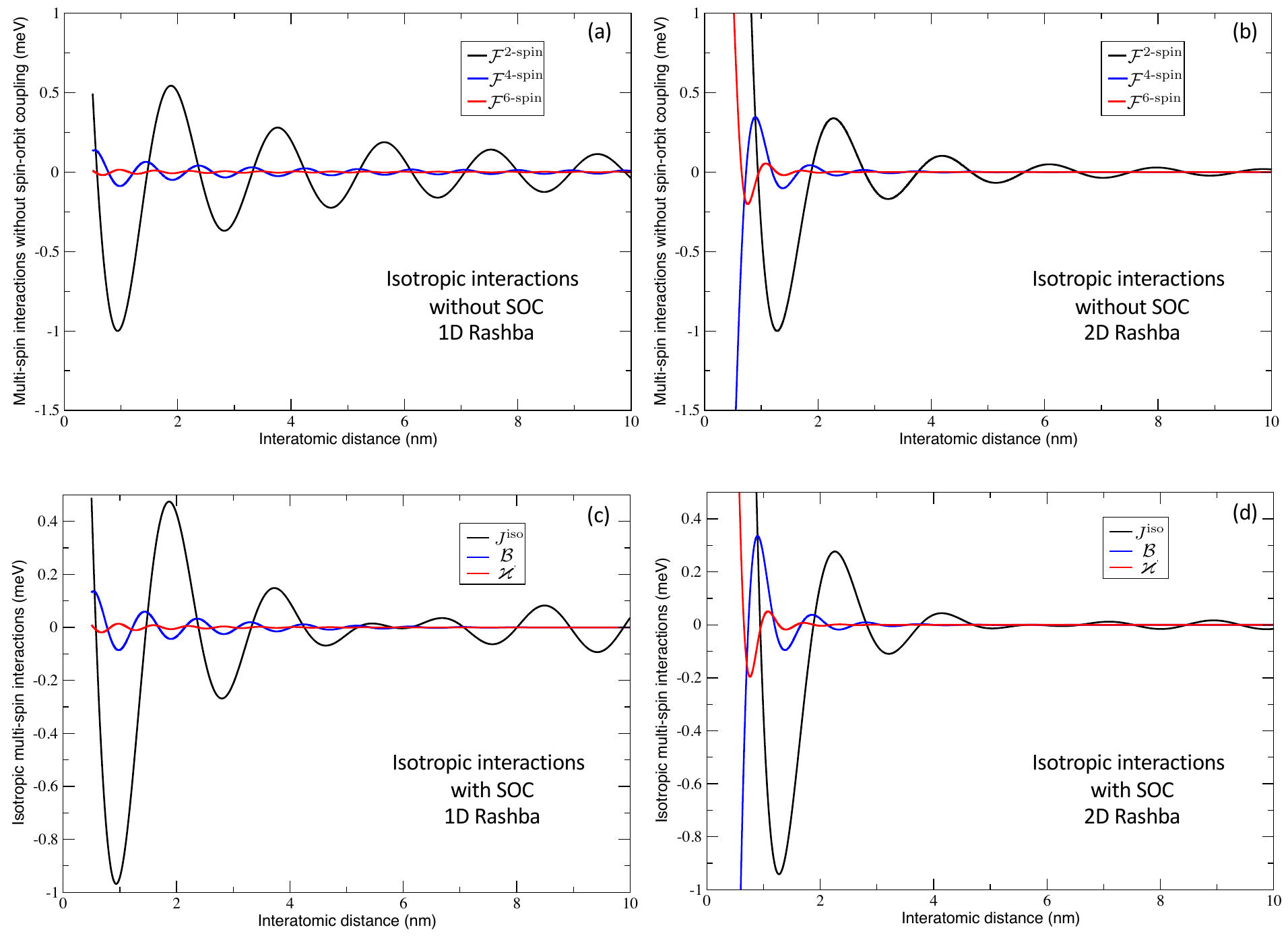}
\caption{Comparison of the isotropic multi-spin interactions without (a-b) and with the Rashba spin-orbit coupling (c-d). The left- (right-) hand-side figures were obtained for the 1D (2D) Rashba model utilizing parameters pertaining to Au(111) surface state~\cite{Walls2007}. The magnetic moments are assumed to be identical and equidistant to facilitate the discussion. The $s-d$ interaction $J^{sd}$ is chosen such that for both the 1D and 2D electron gas, the bilinear magnetic exchange interaction at the first minimum of the oscillatory function  equals -1 meV, favoring an antiferromagnetic coupling. This is motivated by measurements of the long-range magnetic interactions of Co adatoms on Pt(111) surface~\cite{Zhou2010}. }\label{fig:Comparison_isotropic_MEI}
\end{figure*}

\begin{figure*}
\centering
\includegraphics[width=\columnwidth]{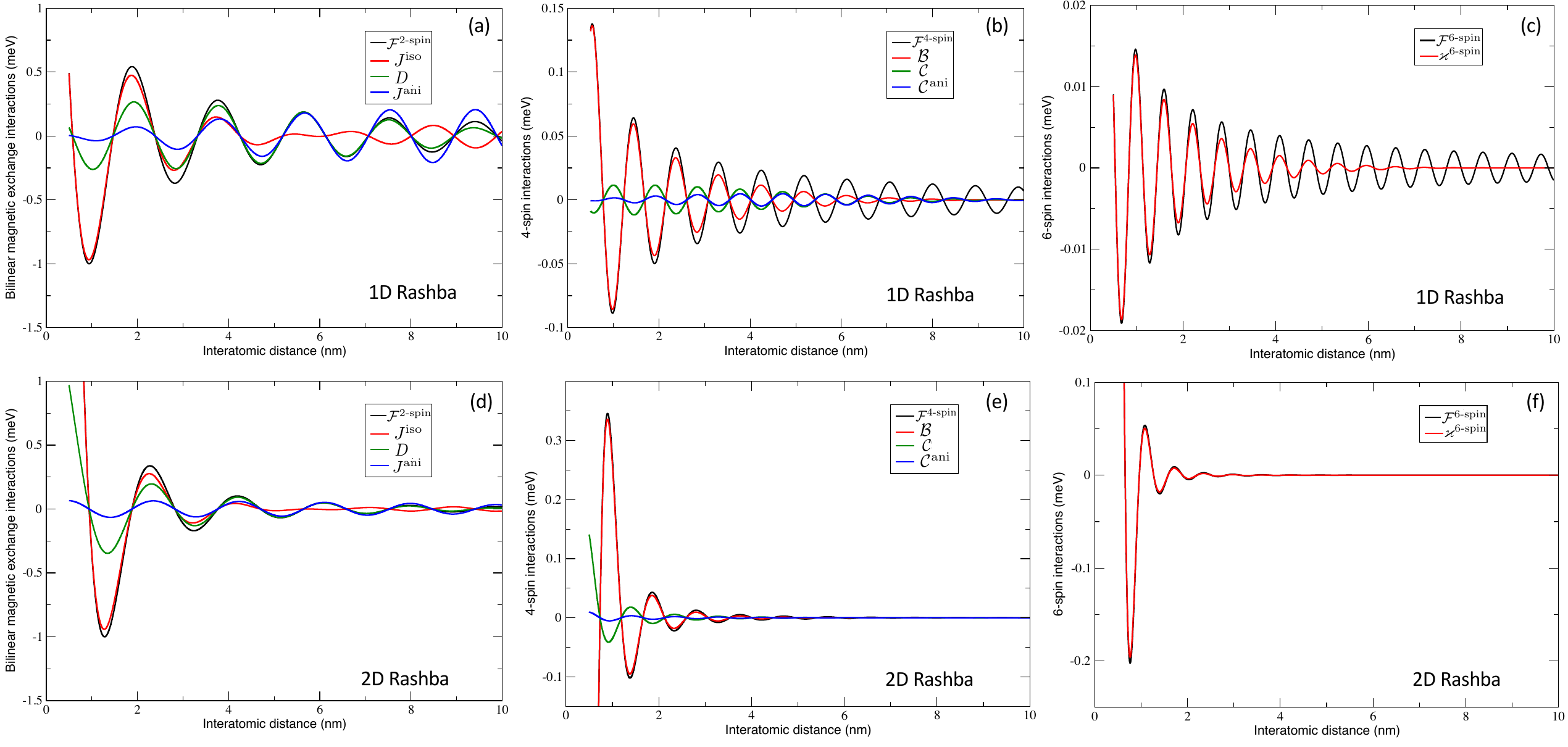}
\caption{Bilinear (a-d), four-spin (b-e) and six-spin (d-f) long-range magnetic exchange interations as function of interatomic distances between identical magnetic moments assuming the parameters describing Au(111) surface state~\cite{Walls2007}. The upper (lower) figures were obtained for the 1D (2D) Rashba model. While the chiral two-spin interactions are relevant at relatively small or large distances, the chiral four-spin interactions seem more important at short distances with their chirality (sign of interaction) can be opposite. Higher-order magnetic interactions are characterized by a larger number of oscillations, which permits a rich phase diagram for small modification of the interatomic distances.  }\label{fig:Comparison_all_MEI}
\end{figure*}

\begin{figure*}
\centering
\includegraphics[width=\columnwidth]{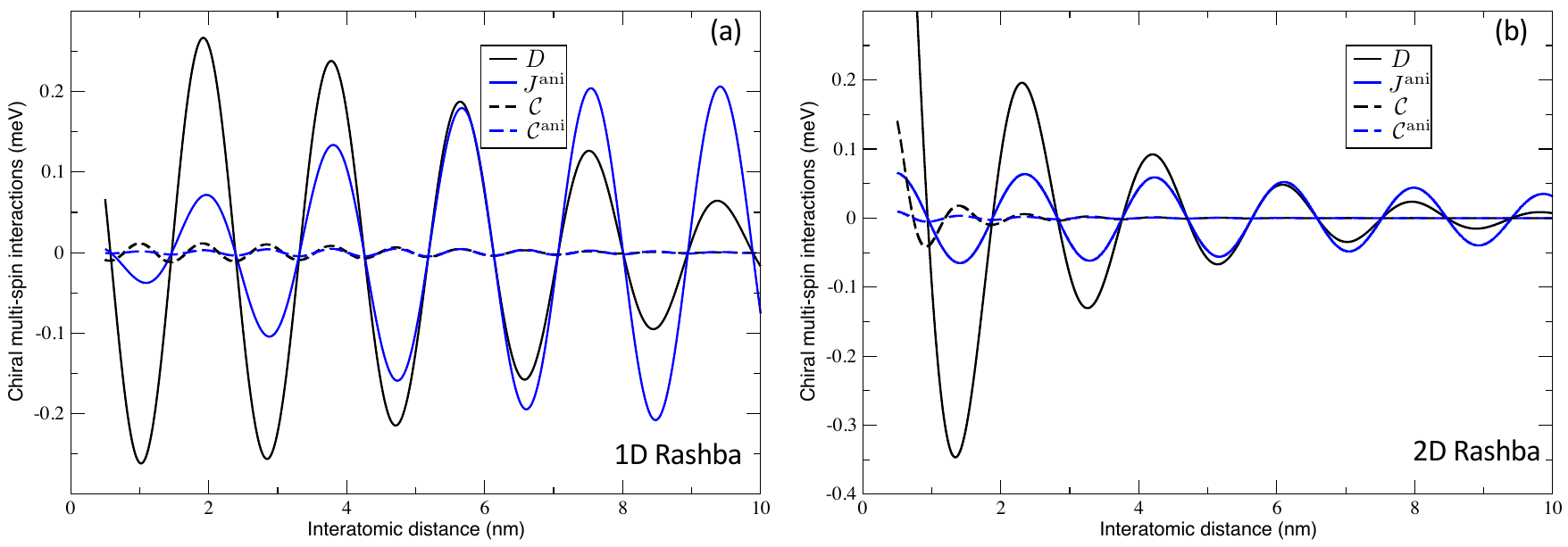}
\caption{Comparison of bilinear and four-spin chiral magnetic exchange interactions. For the 1D Rashba electron gas (a), the bilinear DMI and anisotropic MEI increase in magnitude at large distances where interference effects annihilate the isotropic MEI. The four-spin chiral interactions oscillate with an opposite phase with respect to the bilinear counterparts. For the 2D Rashba electrons, the dephasing is not perfect at relatively short distances, e.g. at 1 nm, which permits to have a chiral four-spin interaction that is much larger than the DMI. Note that the results were obtained with  parameters of the surface state of Au(111)~\cite{Walls2007}.}\label{fig:Comparison_chiral_MEI}
\end{figure*}

\subsection{Numerical evaluation of multi-spin interactions}
I consider the case of Au surfaces characterized by a surface state confined in 2D or 1D with the effective mass $m^* = 0.26\, m_e$, the Fermi wave vector $q_F=0.167~\si{\angstrom^{-1}}$ and $k_R = 0.0135~\si{\angstrom^{-1}}$~\cite{Walls2007,Bouaziz2017}. I address the particular case of equidistant magnetic moments and therefore the MEI are discussed and analysed in terms of the interatomic distance $R$. It is important to keep in mind, that the contribution of the multi-spin interactions to the total energy increases in a factorial fashion when increasing the number of magnetic moments of the plaquettes. Therefore, although the coefficients defining the MEI might be smaller than the bilinear ones, their overall impact on the energy can be prominent. The parameter $J^{sd}$ is chosen such to give the right order of magnitude of the first minimum of the bilinear RKKY interaction, which corresponds to an antiferromagnetic coupling of two magnetic moments. To allow a simple comparison between the 1D and 2D cases, the first largest antiferromagnetic coupling is assumed to be the same  and equal to 1 meV, which is about the value measured for the RKKY interactions between Co adatoms on Pt(111) surface~\cite{Zhou2010}. 

I start by analysing the isotropic magnetic interactions without the Rashba SOC shown in Figs.~\ref{fig:Comparison_isotropic_MEI}(a-b). Note that for the bilinear interactions, a positive (negative) value corresponds to a ferromagnetic (antiferromagnetic) coupling. As expected, the number of oscillations increases with the number of sites involved in the interactions since the wavelength is given by $2\frac{\pi}{Nq_F} = \frac{\lambda_F}{N}$, $N$ being the number of spins involved in the interaction while $R$ is the interatomic distance, which for simplicity is assumed to be identical for all pairs of magnetic moments. This gives rise to situations where the higher order MEI are larger in magnitude than the bilinear ones. As expected, the 2D MEI decay more strongly  than the 1D ones. This advocates for the investigation of the long-range multi-spin interactions in systems with reduced dimensionality.

As illustrated in Figs.~\ref{fig:Comparison_isotropic_MEI}(c-d) (see also Fig.~\ref{fig:Comparison_all_MEI}(d-f) for a better resolution of the six-spin interaction), the inclusion of SOC reduces the amplitude of the isotropic multi-spin interactions with the emergence of a beating effect leading to a vanishing of the interactions accompanied with a phase switch of their oscillations, which is well known for the two-spin interactions~\cite{Bouaziz2017}. This occurs at a length scale defined by the Rashba wave vector, $k_R$, i.e. around 6 nm.

In Figs.~\ref{fig:Comparison_all_MEI}(a-d), the bilinear MEI are shown for the spins embedded respectively in the 1D and 2D Rashba electron gas. Similarly to the isotropic MEI, the amplitude of DMI and $J^\text{ani}$ decreases by increasing the dimension of the electron gas. Making the moments further apart  eventually favors the enhancement of the SO-driven interactions, which contrast strongly with the evanescence of the isotropic interactions around 6 nm.

The four-spin interactions are displayed in Figs.~\ref{fig:Comparison_all_MEI}(b-e). These are initially, i.e. for relatively short interatomic distances, the largest in 2D   but they experience a faster weakening than those of the 1D Rashba gas.  Note that in contrast to the bilinear counterparts, the chiral four-spin interactions oscillate in opposite phase with respect to that of the isotropic ones. 

It is instructive to compare the behavior of the chiral bilinear with that of the equivalent four-spin interactions (see Fig.~\ref{fig:Comparison_chiral_MEI}). Although the amplitude of the four-spin interactions is weaker than that of the bilinear MEI, one notices that at short distances, their sign is opposite and owing to their different wavelengths, there are regions where the chirality behavior is settled by the four-spin interaction. For instance, this occurs at $\approx 1nm$ in the 2D case, where the isotropic MEI is negligible (see Figs.~\ref{fig:Comparison_isotropic_MEI}(b-d).

\section{Conclusion}
In this article, I present a theoretical framework for the evaluation of multi-spin interactions utilizing multiple-scattering theory. This fits methodologies based on the calculation of Green functions, such as the Korringa-Kohn-Rostoker Green functions. I discuss the bilinear, four- and six-spin interactions with a particular focus on isotropic and chiral terms. Then I use this theory for the evaluation of multi-spin interactions for localized spins embedded in one- and two-dimensional electrons described by the Rashba model. Utilizing the RKKY approximation, the latter model offers the possibility of extracting analytically the long-range/asymptotic behavior of isotropic and chiral multi-spin interactions.

Within this approach, there is for each couple of sites $i$ and $j$ a four-spin chiral vector perpendicular to the bond connecting the two sites similarly to their DMI vector. The reported study shows that the 
strong contrast between the 1D and 2D bilinear magnetic exchange interactions survives for higher-order interactions. I recover the power-law decay pertaining to the two-spin magnetic exchange interactions, $\approx q_F^{d-2}R^{-d}$, which I generalize to the $N$-spin case as $\{q_F^{(N-d)}P^{\frac{1}{2}(d-1)}L\}^{-1}  \approx \{q_F^{(N-d)}R^{1+\frac{N}{2}(d-1)}N\}^{-1}$, where $d$ is the dimension of the electron gas mediating the interactions, $L$ the perimeter of the plaquette of $N$ spins while P is the product of interatomic distances and $q_F$ the Fermi wave vector. The 2-, 4- and 6-spin 1D MEI experience a similar same decay, i.e. same power law, with respect to the spatial separation of the spins. In contrast,  the more spins involved in the 2D MEI, the stronger is the decay. Moreover, the dependence with respect to $q_F$ provides a path of engineering the magnitude of the higher-order MEI by tuning the electronic occupation. 

Regarding relativistic effects, obviously the chiral multi-spin interactions cancel out without SOC. The smallest dependence with respect to the latter is linear at short distances. At atomic separations of the order of SOC length scale, beating effects reduce the amplitude of the isotropic MEI giving the opportunity for the chiral interactions to have a strong impact on the magnetic behavior of the investigated systems.  

Numerical results were presented on the basis of  Rashba parameters mimicking surface states residing on Au surfaces. This permits a visual comparison of the various interactions. Their distinct oscillatory behavior  offers the possibility of exploring rich magnetic phase diagrams, which can be strongly altered when modifying the interatomic distances. This is probably  possible  with atomic manipulation based on scanning tunneling microscopy. Additional tuning parameters are the strength of SOC and the electronic occupation, which can be changed  by gating or by modifying the nature of the substrate~\cite{Varykhalov2012,Ishida2014,Bouhassoune2019}. In the future, it would be interesting to extend the current study to the 3D case, either by using a 3D Rashba model~\cite{Feng2019} or by extending the Fert-Levy model~\cite{Fert1980} to the multi-spin interactions.

\section*{Acknowledgements}
I acknowledge fruitful discussions with S. Brinker, M. dos Santos Dias, J. Bouaziz, S. Grytsiuk and S. Bl\"ugel. Funding is provided from the European Research
Council (ERC) under the European Union's Horizon 2020 research and
innovation programme (ERC-consolidator Grant No. 681405 DYNASORE).


\clearpage
\section*{References}
\bibliography{bibliosave}

\providecommand{\newblock}{}
\begin{thebibliography}{100}
\expandafter\ifx\csname url\endcsname\relax
  \def\url#1{{\tt #1}}\fi
\expandafter\ifx\csname urlprefix\endcsname\relax\def\urlprefix{URL }\fi
\providecommand{\eprint}[2][]{\url{#2}}

\bibitem{Parkin1990}
Parkin S~S~P, More N and Roche K~P 1990 {\em Phys. Rev. Lett.\/} {\bf 64}(19)
  2304--2307
  \urlprefix\url{https://link.aps.org/doi/10.1103/PhysRevLett.64.2304}

\bibitem{Fert2017}
Fert A, Reyren N and Cros V 2017 {\em Nature Reviews Materials\/} {\bf 2}
  natrevmats201731
  \urlprefix\url{https://www.nature.com/articles/natrevmats201731}

\bibitem{kanazawa_critical_2016}
Kanazawa N, Nii Y, Zhang X~X, Mishchenko A~S, De~Filippis G, Kagawa F, Iwasa Y,
  Nagaosa N and Tokura Y 2016 {\em Nature Communications\/} {\bf 7} 11622 ISSN
  2041-1723 \urlprefix\url{http://www.nature.com/doifinder/10.1038/ncomms11622}

\bibitem{Bornemann2019}
Bornemann M, Grytsiuk S, Baumeister P~F, dos Santos~Dias M, Zeller R, Lounis S
  and Bl\"ugel S 2019 {\em Journal of Physics: Condensed Matter\/} {\bf 31}
  485801 \urlprefix\url{https://doi.org/10.1088%2F1361-648x%2Fab38a0}

\bibitem{Wang2019}
Wang X~S, Qaiumzadeh A and Brataas A 2019 {\em Phys. Rev. Lett.\/} {\bf
  123}(14) 147203
  \urlprefix\url{https://link.aps.org/doi/10.1103/PhysRevLett.123.147203}

\bibitem{Read2000}
Read N and Green D 2000 {\em Phys. Rev. B\/} {\bf 61}(15) 10267--10297
  \urlprefix\url{https://link.aps.org/doi/10.1103/PhysRevB.61.10267}

\bibitem{Nayak2008}
Nayak C, Simon S~H, Stern A, Freedman M and Das~Sarma S 2008 {\em Rev. Mod.
  Phys.\/} {\bf 80}(3) 1083--1159
  \urlprefix\url{https://link.aps.org/doi/10.1103/RevModPhys.80.1083}

\bibitem{Nadj-Perge2013}
Nadj-Perge S, Drozdov I~K, Bernevig B~A and Yazdani A 2013 {\em Phys. Rev. B\/}
  {\bf 88}(2) 020407
  \urlprefix\url{https://link.aps.org/doi/10.1103/PhysRevB.88.020407}

\bibitem{Klinovaja2013}
Klinovaja J, Stano P, Yazdani A and Loss D 2013 {\em Phys. Rev. Lett.\/} {\bf
  111}(18) 186805
  \urlprefix\url{https://link.aps.org/doi/10.1103/PhysRevLett.111.186805}

\bibitem{Braunecker2013}
Braunecker B and Simon P 2013 {\em Phys. Rev. Lett.\/} {\bf 111}(14) 147202
  \urlprefix\url{https://link.aps.org/doi/10.1103/PhysRevLett.111.147202}

\bibitem{Vazifeh2013}
Vazifeh M~M and Franz M 2013 {\em Phys. Rev. Lett.\/} {\bf 111}(20) 206802
  \urlprefix\url{https://link.aps.org/doi/10.1103/PhysRevLett.111.206802}

\bibitem{Kim2014}
Kim Y, Cheng M, Bauer B, Lutchyn R~M and Das~Sarma S 2014 {\em Phys. Rev. B\/}
  {\bf 90}(6) 060401
  \urlprefix\url{https://link.aps.org/doi/10.1103/PhysRevB.90.060401}

\bibitem{Feldman2017}
Feldman B~E, Randeria M~T, Li J, Jeon S, Xie Y, Wang Z, Drozdov I~K,
  Andrei~Bernevig B and Yazdani A 2017 {\em Nature Physics\/} {\bf 13} 286--291
  \urlprefix\url{https://doi.org/10.1038/nphys3947}

\bibitem{Kim2018}
Kim H, Palacio-Morales A, Posske T, R{\'o}zsa L, Palot{\'a}s K, Szunyogh L,
  Thorwart M and Wiesendanger R 2018 {\em Science Advances\/} {\bf 4}
  (\textit{Preprint}
  \eprint{https://advances.sciencemag.org/content/4/5/eaar5251.full.pdf})
  \urlprefix\url{https://advances.sciencemag.org/content/4/5/eaar5251}

\bibitem{Schneider2020}
Schneider L, Brinker S, Steinbrecher Manuel H, Posske T, dos Santos~Dias M,
  Lounis S, Wiesendanger R and Wiebe J 2020 {\em ArXiv:2002.12294\/}

\bibitem{Fu2008}
Fu L and Kane C~L 2008 {\em Phys. Rev. Lett.\/} {\bf 100}(9) 096407
  \urlprefix\url{https://link.aps.org/doi/10.1103/PhysRevLett.100.096407}

\bibitem{Sau2010}
Sau J~D, Lutchyn R~M, Tewari S and Das~Sarma S 2010 {\em Phys. Rev. Lett.\/}
  {\bf 104}(4) 040502
  \urlprefix\url{https://link.aps.org/doi/10.1103/PhysRevLett.104.040502}

\bibitem{Alicea2010}
Alicea J 2010 {\em Phys. Rev. B\/} {\bf 81}(12) 125318
  \urlprefix\url{https://link.aps.org/doi/10.1103/PhysRevB.81.125318}

\bibitem{Lutchyn2010}
Lutchyn R~M, Sau J~D and Das~Sarma S 2010 {\em Phys. Rev. Lett.\/} {\bf 105}(7)
  077001
  \urlprefix\url{https://link.aps.org/doi/10.1103/PhysRevLett.105.077001}

\bibitem{Oreg2010}
Oreg Y, Refael G and von Oppen F 2010 {\em Phys. Rev. Lett.\/} {\bf 105}(17)
  177002
  \urlprefix\url{https://link.aps.org/doi/10.1103/PhysRevLett.105.177002}

\bibitem{Heisenberg1928}
Heisenberg W 1928 {\em Z. Phys.\/} {\bf 49} 619--636 ISSN 0044-3328

\bibitem{Anderson1959}
Anderson P~W 1959 {\em Phys. Rev.\/} {\bf 115}(1) 2--13

\bibitem{Dzyaloshinskii1958}
Dzyaloshinsky I 1958 {\em J. Phys. Chem. Solids\/} {\bf 4} 241 -- 255
  \urlprefix\url{http://www.sciencedirect.com/science/article/pii/0022369758900763}

\bibitem{Moriya1960}
Moriya T 1960 {\em Phys. Rev.\/} {\bf 120}(1) 91--98
  \urlprefix\url{https://link.aps.org/doi/10.1103/PhysRev.120.91}

\bibitem{vanVleck1937}
van Vleck J~H 1937 {\em Phys. Rev.\/} {\bf 52}(11) 1178--1198

\bibitem{Moriya1953}
Moriya T and Yosida K 1953 {\em Progress of Theoretical Physics\/} {\bf 9}
  663--675 ISSN 0033-068X (\textit{Preprint}
  \eprint{https://academic.oup.com/ptp/article-pdf/9/6/663/5467620/9-6-663.pdf})

\bibitem{Roger1983}
Roger M, Hetherington J~H and Delrieu J~M 1983 {\em Rev. Mod. Phys.\/} {\bf 55}
  1--64

\bibitem{Roger1998}
Roger M, B\"auerle C, Bunkov Y~M, Chen A~S and Godfrin H 1998 {\em Phys. Rev.
  Lett.\/} {\bf 80}(6) 1308--1311

\bibitem{Lounis2010}
Lounis S and Dederichs P~H 2010 {\em Phys. Rev. B\/} {\bf 82}(18) 180404
  \urlprefix\url{https://link.aps.org/doi/10.1103/PhysRevB.82.180404}

\bibitem{Szilva2013}
Szilva A, Costa M, Bergman A, Szunyogh L, Nordstr\"om L and Eriksson O 2013
  {\em Phys. Rev. Lett.\/} {\bf 111}(12) 127204
  \urlprefix\url{https://link.aps.org/doi/10.1103/PhysRevLett.111.127204}

\bibitem{Al-Zubi2011}
Al-Zubi A, Bihlmayer G and Bl\"ugel S 2011 {\em Phys. Status Solidi B\/} {\bf
  248} 2242--2247
  \urlprefix\url{https://onlinelibrary.wiley.com/doi/abs/10.1002/pssb.201147090}

\bibitem{Kroenlein2018}
Kr\"onlein A, Schmitt M, Hoffmann M, Kemmer J, Seubert N, Vogt M, K\"uspert J,
  B\"ohme M, Alonazi B, K\"ugel J, Albrithen H~A, Bode M, Bihlmayer G and
  Bl\"ugel S 2018 {\em Phys. Rev. Lett.\/} {\bf 120}(20) 207202

\bibitem{Romming2018}
Romming N, Pralow H, Kubetzka A, Hoffmann M, von Malottki S, Meyer S, Dup\'e B,
  Wiesendanger R, von Bergmann K and Heinze S 2018 {\em Phys. Rev. Lett.\/}
  {\bf 120}(20) 207201

\bibitem{Kittel1960}
Kittel C 1960 {\em Phys. Rev.\/} {\bf 120}(2) 335--342

\bibitem{Harris1963}
Harris E~A and Owen J 1963 {\em Phys. Rev. Lett.\/} {\bf 11}(1) 9--10

\bibitem{Huang1964}
Huang N~L and Orbach R 1964 {\em Phys. Rev. Lett.\/} {\bf 12}(11) 275--276

\bibitem{Brinker2019}
Brinker S, Dias M~d~S and Lounis S 2019 {\em New Journal of Physics\/} {\bf 21}
  083015 ISSN 1367-2630

\bibitem{Uryu1965}
Ury\^u N and Friedberg S~A 1965 {\em Phys. Rev.\/} {\bf 140}(5A) A1803--A1811

\bibitem{Iwashita1974}
Iwashita T and Uryû N 1974 {\em Journal of the Physical Society of Japan\/}
  {\bf 36} 48--54 (\textit{Preprint}
  \eprint{https://doi.org/10.1143/JPSJ.36.48})

\bibitem{Hoffmann2020}
Hoffmann M and Bl\"ugel S 2020 {\em Phys. Rev. B\/} {\bf 101} 024418

\bibitem{Laszloffy2019}
L{\'a}szl{\'o}ffy A, R{\'o}zsa L, Palot{\'a}s K, Udvardi L and Szunyogh L 2019
  {\em Physical Review B\/} {\bf 99} 184430

\bibitem{Grytsiuk2020}
Grytsiuk S, Hanke J~P, Hoffmann M, Bouaziz J, Gomonay O, Bihlmayer G, Lounis S,
  Mokrousov Y and Bl{\"u}gel S 2020 {\em Nature Communications\/} {\bf 11} 1--7
  ISSN 2041-1723

\bibitem{Brinker2020}
Brinker S, dos Santos~Dias M and Lounis S 2020 {\em Phys. Rev. Research\/} {\bf
  2}(3) 033240
  \urlprefix\url{https://link.aps.org/doi/10.1103/PhysRevResearch.2.033240}

\bibitem{Hoffmann2015}
Hoffmann M, Weischenberg J, Dup\'e B, Freimuth F, Ferriani P, Mokrousov Y and
  Heinze S 2015 {\em Phys. Rev. B\/} {\bf 92}(2) 020401
  \urlprefix\url{https://link.aps.org/doi/10.1103/PhysRevB.92.020401}

\bibitem{Dias2016}
dos Santos~Dias M, Bouaziz J, Bouhassoune M, Bl\"ugel S and Lounis S 2016 {\em
  Nature Commun.\/} {\bf 7} 13613

\bibitem{Hanke2016}
Hanke J~P, Freimuth F, Nandy A~K, Zhang H, Bl\"ugel S and Mokrousov Y 2016 {\em
  Phys. Rev. B\/} {\bf 94}(12) 121114
  \urlprefix\url{https://link.aps.org/doi/10.1103/PhysRevB.94.121114}

\bibitem{Dias2017}
dos Santos~Dias M and Lounis S 2017 {\em Spintronics X\/} {\bf 10357} 136 --
  152 \urlprefix\url{https://doi.org/10.1117/12.2275305}

\bibitem{Ruderman1954}
Ruderman M~A and Kittel C 1954 {\em Phys. Rev.\/} {\bf 96}(1) 99--102
  \urlprefix\url{https://link.aps.org/doi/10.1103/PhysRev.96.99}

\bibitem{Kasuya1956}
Kasuya T 1956 {\em Progress of Theoretical Physics\/} {\bf 16} 45--57 ISSN
  0033-068X \urlprefix\url{https://doi.org/10.1143/PTP.16.45}

\bibitem{Yosida1957}
Yosida K 1957 {\em Phys. Rev.\/} {\bf 106}(5) 893--898
  \urlprefix\url{https://link.aps.org/doi/10.1103/PhysRev.106.893}

\bibitem{Gruenberg1986}
Gr\"unberg P, Schreiber R, Pang Y, Brodsky M~B and Sowers H 1986 {\em Phys.
  Rev. Lett.\/} {\bf 57}(19) 2442--2445
  \urlprefix\url{https://link.aps.org/doi/10.1103/PhysRevLett.57.2442}

\bibitem{Carbone1987}
Carbone C and Alvarado S~F 1987 {\em Phys. Rev. B\/} {\bf 36}(4) 2433--2435
  \urlprefix\url{https://link.aps.org/doi/10.1103/PhysRevB.36.2433}

\bibitem{Bruno1991}
Bruno P and Chappert C 1991 {\em Phys. Rev. Lett.\/} {\bf 67}(12) 1602--1605
  \urlprefix\url{https://link.aps.org/doi/10.1103/PhysRevLett.67.1602}

\bibitem{Batista2016}
Batista C~D, Lin S~Z, Hayami S and Kamiya Y 2016 {\em Reports on Progress in
  Physics\/} {\bf 79} 084504
  \urlprefix\url{http://stacks.iop.org/0034-4885/79/i=8/a=084504}

\bibitem{Ozawa2017}
Ozawa R, Hayami S and Motome Y 2017 {\em Phys. Rev. Lett.\/} {\bf 118}(14)
  147205

\bibitem{Hayami2017}
Hayami S, Ozawa R and Motome Y 2017 {\em Phys. Rev. B\/} {\bf 95}(22) 224424

\bibitem{Okumura2020}
Okumura S, Hayami S, Kato Y and Motome Y 2020 {\em Phys. Rev. B\/} {\bf
  101}(14) 144416
  \urlprefix\url{https://link.aps.org/doi/10.1103/PhysRevB.101.144416}

\bibitem{Heinze2011}
Heinze S, von Bergmann K, Menzel M, Brede J, Kubetzka A, Wiesendanger R,
  Bihlmayer G and Bl{\"u}gel S 2011 {\em Nat. Phys.\/} {\bf 7} 713 EP --
  \urlprefix\url{https://doi.org/10.1038/nphys2045}

\bibitem{Tanigaki2015}
Tanigaki T, Shibata K, Kanazawa N, Yu X, Onose Y, Park H~S, Shindo D and Tokura
  Y 2015 {\em Nano Letters\/} {\bf 15} 5438--5442 ISSN 1530-6984

\bibitem{Takagieaau2018}
Takagi R, White J~S, Hayami S, Arita R, Honecker D, R{\o}nnow H~M, Tokura Y and
  Seki S 2018 {\em Science Advances\/} {\bf 4} eaau3402

\bibitem{Fujishiro2019}
Fujishiro Y, Kanazawa N, Nakajima T, Yu X~Z, Ohishi K, Kawamura Y, Kakurai K,
  Arima T, Mitamura H, Miyake A, Akiba K, Tokunaga M, Matsuo A, Kindo K,
  Koretsune T, Arita R and Tokura Y 2019 {\em Nature Communications\/} {\bf 10}
  1059 ISSN 2041-1723

\bibitem{Kurumaji2019}
Kurumaji T, Nakajima T, Hirschberger M, Kikkawa A, Yamasaki Y, Sagayama H,
  Nakao H, Taguchi Y, Arima T~h and Tokura Y 2019 {\em Science\/} {\bf 365}
  914--918 ISSN 0036-8075
  \urlprefix\url{https://science.sciencemag.org/content/365/6456/914}

\bibitem{Khanh2020}
Khanh N~D, Nakajima T, Yu X~Z, Gao S, Shibata K, Hirschberger M, Yamasaki Y,
  Sagayama H, Nakao H, Peng L~C, Nakajima K, Takagi R, Arima T, Tokura Y and
  Seki S 2020 Nanometric square skyrmion lattice in a centrosymmetric
  tetragonal magnet (\textit{Preprint} \eprint{2003.00626})

\bibitem{Braunecker2009}
Braunecker B, Simon P and Loss D 2009 {\em Phys. Rev. B\/} {\bf 80}(16) 165119
  \urlprefix\url{https://link.aps.org/doi/10.1103/PhysRevB.80.165119}

\bibitem{Meng2013}
Meng T and Loss D 2013 {\em Phys. Rev. B\/} {\bf 87}(23) 235427
  \urlprefix\url{https://link.aps.org/doi/10.1103/PhysRevB.87.235427}

\bibitem{Friedel1952}
Friedel J 1952 {\em Philosophical Magazine Series\/} {\bf 7} 153

\bibitem{Walls2007}
Walls J~D and Heller E~J 2007 {\em Nano Letters\/} {\bf 7} 3377--3382
  \urlprefix\url{https://doi.org/10.1021/nl071711z}

\bibitem{Lounis2012}
Lounis S, Bringer A and Bl\"ugel S 2012 {\em Phys. Rev. Lett.\/} {\bf 108}(20)
  207202
  \urlprefix\url{https://link.aps.org/doi/10.1103/PhysRevLett.108.207202}

\bibitem{Smith1976}
Smith D 1976 {\em Journal of Magnetism and Magnetic Materials\/} {\bf 1} 214 --
  225 ISSN 0304-8853
  \urlprefix\url{http://www.sciencedirect.com/science/article/pii/030488537690069X}

\bibitem{Fert1980}
Fert A and Levy P~M 1980 {\em Phys. Rev. Lett.\/} {\bf 44}(23) 1538--1541
  \urlprefix\url{https://link.aps.org/doi/10.1103/PhysRevLett.44.1538}

\bibitem{Imamura2004}
Imamura H, Bruno P and Utsumi Y 2004 {\em Phys. Rev. B\/} {\bf 69}(12) 121303
  \urlprefix\url{https://link.aps.org/doi/10.1103/PhysRevB.69.121303}

\bibitem{Khajetoorians2016}
Khajetoorians A~A, Steinbrecher M, Ternes M, Bouhassoune M, dos Santos~Dias M,
  Lounis S, Wiebe J and Wiesendanger R 2016 {\em Nature Communications\/} {\bf
  7} 10620 \urlprefix\url{https://doi.org/10.1038/ncomms10620}

\bibitem{Bouaziz2017}
Bouaziz J, dos Santos~Dias M, Ziane A, Benakki M, Blügel S and Lounis S 2017
  {\em New Journal of Physics\/} {\bf 19} 023010

\bibitem{Han2019}
Han D~S, Lee K, Hanke J~P, Mokrousov Y, Kim K~W, Yoo W, van Hees Y~L~W, Kim
  T~W, Lavrijsen R, You C~Y, Swagten H~J~M, Jung M~H and Kl{\"a}ui M 2019 {\em
  Nature Materials\/} {\bf 18} 703--708
  \urlprefix\url{https://doi.org/10.1038/s41563-019-0370-z}

\bibitem{Schmitt2019}
Schmitt M, Moras P, Bihlmayer G, Cotsakis R, Vogt M, Kemmer J, Belabbes A,
  Sheverdyaeva P~M, Kundu A~K, Carbone C, Bl{\"u}gel S and Bode M 2019 {\em
  Nature Communications\/} {\bf 10} 2610
  \urlprefix\url{https://doi.org/10.1038/s41467-019-10515-3}

\bibitem{Khajetoorians2013}
Khajetoorians A~A, Schlenk T, Schweflinghaus B, dos Santos~Dias M, Steinbrecher
  M, Bouhassoune M, Lounis S, Wiebe J and Wiesendanger R 2013 {\em Phys. Rev.
  Lett.\/} {\bf 111}(15) 157204
  \urlprefix\url{https://link.aps.org/doi/10.1103/PhysRevLett.111.157204}

\bibitem{Bouhassoune2016}
Bouhassoune M, Dias M~d~S, Zimmermann B, Dederichs P~H and Lounis S 2016 {\em
  Phys. Rev. B\/} {\bf 94}(12) 125402
  \urlprefix\url{https://link.aps.org/doi/10.1103/PhysRevB.94.125402}

\bibitem{Hermenau2019}
Hermenau J, Brinker S, Marciani M, Steinbrecher M, dos Santos~Dias M,
  Wiesendanger R, Lounis S and Wiebe J 2019 {\em Nat. Commun.\/} {\bf 10} 2565
  ISSN 2041-1723

\bibitem{Rashba1960}
Rashba E~I 1960 {\em Sov. Phys. Solid State\/} {\bf 2} 1109

\bibitem{Bychkov1984}
Bychkov Y~A and Rashba E~I 1984 {\em J. Phys. C: Solid State Phys.\/} {\bf 17}
  6039

\bibitem{Reinert2003}
Reinert F 2003 {\em Journal of Physics: Condensed Matter\/} {\bf 15} S693--S705
  \urlprefix\url{https://doi.org/10.1088%2F0953-8984%2F15%2F5%2F321}

\bibitem{Mugarza2001}
Mugarza A, Mascaraque A, P\'erez-Dieste V, Repain V, Rousset S, Garc\'{\i}a~de
  Abajo F~J and Ortega J~E 2001 {\em Phys. Rev. Lett.\/} {\bf 87}(10) 107601
  \urlprefix\url{https://link.aps.org/doi/10.1103/PhysRevLett.87.107601}

\bibitem{Ortega2002}
Ortega J~E, Mugarza A, Repain V, Rousset S, P\'erez-Dieste V and Mascaraque A
  2002 {\em Phys. Rev. B\/} {\bf 65}(16) 165413
  \urlprefix\url{https://link.aps.org/doi/10.1103/PhysRevB.65.165413}

\bibitem{Wahl2007}
Wahl P, Simon P, Diekh\"oner L, Stepanyuk V~S, Bruno P, Schneider M~A and Kern
  K 2007 {\em Phys. Rev. Lett.\/} {\bf 98}(5) 056601
  \urlprefix\url{https://link.aps.org/doi/10.1103/PhysRevLett.98.056601}

\bibitem{Meier:2008}
Meier F, Zhou L, Wiebe J and Wiesendanger R 2008 {\em Science\/} {\bf 320}
  82--86

\bibitem{Zhou2010}
Zhou L, Wiebe J, Lounis S, Vedmedenko E, Meier F, Bl{\"u}gel S, Dederichs P~H
  and Wiesendanger R 2010 {\em Nature Physics\/} {\bf 6} 187

\bibitem{Meier2011}
Meier F, Lounis S, Wiebe J, Zhou L, Heers S, Mavropoulos P, Dederichs P~H,
  Bl\"ugel S and Wiesendanger R 2011 {\em Phys. Rev. B\/} {\bf 83}(7) 075407
  \urlprefix\url{https://link.aps.org/doi/10.1103/PhysRevB.83.075407}

\bibitem{Khajetoorians2012}
Khajetoorians A~A, Wiebe J, Chilian B, Lounis S, Bl{\"u}gel S and Wiesendanger
  R 2012 {\em Nature Physics\/} {\bf 8} 497

\bibitem{Prueser2014}
Pr{\"u}ser H, Dargel P~E, Bouhassoune M, Ulbrich R~G, Pruschke T, Lounis S and
  Wenderoth M 2014 {\em Nature Communications\/} {\bf 5} 5417
  \urlprefix\url{https://doi.org/10.1038/ncomms6417}

\bibitem{Liechtenstein1987}
Liechtenstein A~I, Katsnelson M, Antropov V and Gubanov V 1987 {\em J. Magn.
  Magn. Mater.\/} {\bf 67} 65--74
  \urlprefix\url{http://www.sciencedirect.com/science/article/pii/0304885387907219}

\bibitem{Udvardi2003}
Udvardi L, Szunyogh L, Palot{\'a}s K and Weinberger P 2003 {\em Phys. Rev. B\/}
  {\bf 68} 104436
  \urlprefix\url{http://prb.aps.org/abstract/PRB/v68/i10/e104436}

\bibitem{Ebert2009}
Ebert H and Mankovsky S 2009 {\em Phys. Rev. B\/} {\bf 79} 045209
  \urlprefix\url{http://prb.aps.org/abstract/PRB/v79/i4/e045209}

\bibitem{Heine1980}
Heine V 1980 Electronic structure from the point of view of the local atomic
  environment ({\em Solid State Physics\/} vol~35) ed Ehrenreich H, Seitz F and
  Turnbull D (Academic Press) pp 1 -- 127
  \urlprefix\url{http://www.sciencedirect.com/science/article/pii/S0081194708605032}

\bibitem{Oswald1985}
Oswald A, Zeller R, Braspenning P~J and Dederichs P~H 1985 {\em Journal of
  Physics F: Metal Physics\/} {\bf 15} 193--212
  \urlprefix\url{https://doi.org/10.1088%2F0305-4608%2F15%2F1%2F021}

\bibitem{Papanikolaou2002}
Papanikolaou N, Zeller R and Dederichs P~H 2002 {\em J. Phys.: Condens.
  Matter\/} {\bf 14} 2799--2823
  \urlprefix\url{https://doi.org/10.1088%2F0953-8984%2F14%2F11%2F304}

\bibitem{Lloyd1972}
Lloyd P and Smith P 1972 {\em Advances in Physics\/} {\bf 21} 69--142
  (\textit{Preprint} \eprint{https://doi.org/10.1080/00018737200101268})
  \urlprefix\url{https://doi.org/10.1080/00018737200101268}

\bibitem{Drittler1989}
Drittler B, Weinert M, Zeller R and Dederichs P~H 1989 {\em Phys. Rev. B\/}
  {\bf 39}(2) 930--939
  \urlprefix\url{https://link.aps.org/doi/10.1103/PhysRevB.39.930}

\bibitem{Cardias2020}
{Cardias} R, {Bergman} A, {Szilva} A, {Kvashnin} Y~O, {Fransson} J, {Klautau}
  A~B, {Eriksson} O and {Nordstr{\"o}m} L 2020 {\em arXiv e-prints\/}
  arXiv:2003.04680 (\textit{Preprint} \eprint{2003.04680})

\bibitem{Yafet1987}
Yafet Y 1987 {\em Phys. Rev. B\/} {\bf 36}(7) 3948--3949
  \urlprefix\url{https://link.aps.org/doi/10.1103/PhysRevB.36.3948}

\bibitem{Litvinov1998}
Litvinov V~I and Dugaev V~K 1998 {\em Phys. Rev. B\/} {\bf 58}(7) 3584--3585
  \urlprefix\url{https://link.aps.org/doi/10.1103/PhysRevB.58.3584}

\bibitem{Varykhalov2012}
Varykhalov A, Marchenko D, Scholz M~R, Rienks E~D~L, Kim T~K, Bihlmayer G,
  S\'anchez-Barriga J and Rader O 2012 {\em Phys. Rev. Lett.\/} {\bf 108}(6)
  066804
  \urlprefix\url{https://link.aps.org/doi/10.1103/PhysRevLett.108.066804}

\bibitem{Ishida2014}
Ishida H 2014 {\em Phys. Rev. B\/} {\bf 90}(23) 235422
  \urlprefix\url{https://link.aps.org/doi/10.1103/PhysRevB.90.235422}

\bibitem{Bouhassoune2019}
Bouhassoune M, Fernandes I~L, Blügel S and Lounis S 2019 {\em New Journal of
  Physics\/} {\bf 21} 063015
  \urlprefix\url{https://doi.org/10.1088%2F1367-2630%2Fab2487}

\bibitem{Feng2019}
Feng Y, Jiang Q, Feng B, Yang M, Xu T, Liu W, Yang X, Arita M, Schwier E~F,
  Shimada K, Jeschke H~O, Thomale R, Shi Y, Wu X, Xiao S, Qiao S and He S 2019
  {\em Nature Communications\/} {\bf 10} 4765
  \urlprefix\url{https://doi.org/10.1038/s41467-019-12805-2}

\end{thebibliography}

\end{document}